\documentclass[11pt]{article}
\usepackage[T1]{fontenc}
\usepackage{mathtools}
\usepackage{jcappub}
\usepackage{graphicx}
\usepackage{color}
\usepackage{amsfonts,amssymb,amsmath,dsfont}
\usepackage{hyperref}
\usepackage{slashed}
\usepackage{braket}
\usepackage[framemethod=TikZ]{mdframed}
\usepackage[normalem]{ulem}
\usepackage{comment}
\usepackage{siunitx}
\usepackage{multirow}
\usepackage{subcaption}
\usepackage{fontawesome}

\usepackage{placeins}
\usepackage{afterpage}
\usepackage{xcolor}
\usepackage{lipsum}
\usepackage{pifont}% http://ctan.org/pkg/pifont
\renewcommand{\thefootnote}{\fnsymbol{footnote}}

\definecolor{ForestGreen}{rgb}{0.13, 0.55, 0.13}
\definecolor{airforceblue}{rgb}{0.36, 0.54, 0.66}
\definecolor{orange}{rgb}{1.0, 0.5, 0.0}

\mdfdefinestyle{MyFrame}{%
    linecolor=black,
    outerlinewidth=0.5pt,
    roundcorner=10pt,
    innertopmargin=\baselineskip,
    innerbottommargin=\baselineskip,
    innerrightmargin=20pt,
    innerleftmargin=20pt,
    backgroundcolor=gray!08!white}

\newcommand{\rhat}{\hat{r}}

\newcommand{\btheta}{\boldsymbol{\theta}}
\newcommand{\bvartheta}{\boldsymbol{\vartheta}}
\newcommand{\boldeta}{\boldsymbol{\eta}}

\newcommand{\bx}{\boldsymbol{x}}

\newcommand{\eg}{\emph{e.g.}}
\newcommand{\ie}{\emph{i.e.}}

\newcommand{\bs}{\boldsymbol{s}}

\newcommand{\mathbbm}[1]{\text{\usefont{U}{bbm}{m}{n}#1}}
\newcommand{\indicator}{\mathbbm{1}}

\newcommand{\GammaPi}{\Gamma_{\pi}}

\newcommand{\swyft}{\texttt{swyft}}
\newcommand{\hillipop}{\texttt{HiLLiPoP}}
\newcommand{\class}{\texttt{CLASS}}
\newcommand{\moped}{\textsc{moped}}

\newcommand{\MCMC}{\textsc{mcmc}}
\newcommand{\SBI}{\textsc{sbi}}
\newcommand{\ABC}{\textsc{abc}}
\newcommand{\TMNRE}{\textsc{tmnre}} % if you want space after this, please type `\TMNRE\` exactly. Thanks!
\newcommand{\MNRE}{\textsc{mnre}}
\newcommand{\pydelfi}{\textsc{pydelfi}}
\newcommand{\delfi}{\textsc{delfi}}

\newcommand{\NLE}{\textsc{nle}}
\newcommand{\NPE}{\textsc{npe}}
\newcommand{\NRE}{\textsc{nre}}

\newcommand{\SNLE}{\textsc{snle}}

\DeclareMathOperator*{\argmin}{arg\,min}
\DeclareMathOperator{\supp}{supp}

\DeclareMathOperator{\supalt}{supp_{\tilde{\epsilon}}}

\title{Fast and Credible Likelihood-Free Cosmology with Truncated Marginal Neural Ratio Estimation}
\author[a, *]{Alex Cole \note{Corresponding author.}}
\author[a]{Benjamin K. Miller}
\author[a]{Samuel J. Witte}
\author[b,c]{Maxwell X. Cai}
\author[d]{Meiert W.\ Grootes}
\author[d]{Francesco Nattino}
\author[a, *]{Christoph Weniger}

\affiliation[a]{Gravitation Astroparticle Physics Amsterdam (GRAPPA), Institute for Theoretical Physics Amsterdam and Delta Institute for Theoretical Physics, University of Amsterdam, Science Park 904, 1098 XH Amsterdam, The Netherlands}
\affiliation[b]{SURF Cooperative, Science Park 140, 1098 XG Amsterdam}
\affiliation[c]{Leiden Observatory, Leiden University, Niels Bohrweg 2, 2300 RA Leiden, The Netherlands}
\affiliation[d]{Netherlands eScience Center, Science Park 140, 1098 XG Amsterdam, The Netherlands}

\emailAdd{a.e.cole2@uva.nl}
\emailAdd{c.weniger@uva.nl}

\abstract{Sampling-based inference techniques are central to modern cosmological data analysis; these methods, however, scale poorly with dimensionality and typically require approximate or intractable likelihoods.
    In this paper we describe how Truncated Marginal Neural Ratio Estimation (\TMNRE) (a new approach in so-called
    simulation-based inference) naturally evades these issues, improving the $(i)$ efficiency, $(ii)$ scalability, and $(iii)$ trustworthiness of the inference. Using measurements of the Cosmic Microwave Background (CMB), we show that \TMNRE\  can achieve converged posteriors using orders of magnitude fewer simulator calls than conventional Markov Chain Monte Carlo (\MCMC) methods. Remarkably, in these examples the required number of samples is effectively independent of the number of nuisance parameters. In addition, a property called \emph{local amortization} allows the performance of rigorous statistical consistency checks that are not accessible to sampling-based methods. \TMNRE\ promises to become a powerful tool for cosmological data analysis, particularly in the context of extended cosmologies, where the timescale required for conventional sampling-based inference methods to converge can greatly exceed that of simple cosmological models such as $\Lambda$CDM. To perform these computations, we use an implementation of \TMNRE\ via the open-source code \swyft.\footnote[3]{\swyft\ is available at \href{https://github.com/undark-lab/swyft}{https://github.com/undark-lab/swyft}. Demonstration on cosmological simulators used in this paper is available at \href{https://github.com/a-e-cole/swyft-CMB}{https://github.com/a-e-cole/swyft-CMB}.}}

\begin{document}
\maketitle

\def\thefootnote{\arabic{footnote}}
\setcounter{footnote}{0}

\section{Introduction}

An impending flood of high-quality data from cosmological surveys promises to revolutionize  our understanding of the universe (see \eg\ \cite{DiValentino:2020vhf} for a review of upcoming surveys).
The collection of this data alone, however, is not sufficient to perform fundamental science; in order to extract cosmological information one must also ``solve the inverse problem,'' \ie\ given a set of observations, one must deduce the set of cosmological models that may have produced them. 

Inferring model parameters from complex observational data is one of the grand challenges of modern cosmology. Part of the difficulty arises from the fact that classical methods rely on evaluating the likelihood of data given model parameters.
Often only an approximation of this likelihood, representing a fraction of the data's information content (\eg\ rather than the entire spatial distribution of galaxies, only the power spectrum of this distribution at large scales), is theoretically understood. In addition, these approaches require sampling the \emph{full joint posterior}, so that time required to achieve convergence scales poorly with the dimensionality of the parameter space. This latter point is particularly troublesome for problems  with large numbers of nuisance parameters, and arises independently of whether the likelihood is known.

Novel approaches in the field of simulation-based inference (\SBI)\footnote{\SBI\ is closely connected to Approximate Bayesian Computation \cite{first_abc, second_abc}, which refers to a class of computational methods used to estimate posterior distributions, and has been developed since the 1980s \cite{diggle1984monte}.} are starting to overcome these obstacles (see \cite{Cranmer2020} for a recent review).
\SBI\ exploits the fact that data generated by a stochastic simulator can be regarded as sampling an exact (implicit) likelihood, even when this function is not explicitly known.
The field has recently seen accelerated progress thanks to deep learning methods \cite{Hermans2019, Cranmer2020, papamakarios2019sequential, greenberg2019automatic,Alsing:2019xrx}, and has been applied to a wide array of problems in cosmology and astrophysics, from cosmological density fields to gravitational waves \cite{Zhao:2021ddh,Makinen:2021nly,Villaescusa-Navarro:2021pkb,Villaescusa-Navarro:2021cni,dax2021real, delaunoy2020lightning}.
In order to improve the scalability with the dimensionality of the parameter space, it is important to note that in many cases the full joint posterior is overly informative. Typically, one is only interested in a subset of 1- and 2-dimensional marginal posteriors; for example, in cosmology, one may be interested in the cosmological parameters but not the large associated list of nuisance parameters parameterizing \eg\ foregrounds and instrumental systematics. By directly targeting low-dimensional posteriors we can alleviate sampling problems associated with high dimensionality, and thus an \textit{arbitrarily} large number of nuisance parameters can be included without additional computational overhead.\footnote{Note that if all parameters contribute equal variance to the data, the implicit data distribution becomes exceedingly noisy. When we refer to scaling to arbitrary number of parameters, the data variance is implicitly held fixed. This limit remains a challenge for sampling-based methods, but is tractable in our proposal.}

In this paper, we describe and apply a new algorithm in simulation-based inference called \emph{Truncated Marginal Neural Ratio Estimation} (\TMNRE)  \cite{Miller:2020hua,Miller:2021hys}.
\TMNRE\ uses neural networks to directly learn \emph{marginal} likelihood-to-evidence ratios in a sequence of rounds. 
After each round, evaluations of the network are used to truncate the prior, so that the parameter region of interest is targeted. In other words, simulations are evaluated in regions of parameter space where the posterior for a given observation is expected to have support. As presented in \cite{Miller:2020hua,Miller:2021hys}, \TMNRE\ has a number of major advantages over classical sampling-based inference techniques (as well as some other \SBI\ algorithms); these advantages can be broadly categorized as
\begin{enumerate}
    \item {\bf Efficiency:} Learning 1- and 2-dimensional marginal posteriors is far simpler than trying to sample the full joint posterior. As such, \TMNRE\ naturally outperforms conventional sampling methods like \MCMC, typically requiring orders of magnitude fewer simulator calls.  In addition, one can store and reuse simulations should one wish to change the priors, the observations, or the inference network. For example, we show in Sec.\ \ref{sec:CMB+BAO} that after having trained a network to infer the posteriors of cosmological parameters using the CMB, one can include BAOs (baryonic acoustic oscillations) to learn corresponding posteriors at {\emph{no additional simulation cost}}.
    \item {\bf Scalability:} As previously mentioned, sampling-based inference approaches scale poorly with the dimensionality of the parameter space. In most problems of interest, the dimensionality can be greatly augmented due to a large list of nuisance parameters, the posteriors of which are typically not of interest. \TMNRE\ offers the possibility of learning the posteriors only for the parameter subset of interest, and thus is easily scalable to high dimensional problems. We demonstrate the excellent scaling behavior of \TMNRE\ on a ``realistic'' Planck simulator in Sec.\ \ref{sec:hillipop}.
    \item {\bf Testability:} Novel approaches to inference, in particular those rooted in machine learning, are often met with a  level of skepticism.
    While in some cases caution is justified,
    the reliability of many \SBI\ algorithms (including \TMNRE) can be explicitly tested, which is \textit{not} easily possible for likelihood-based methods.
    In the context of \TMNRE, thanks to so-called \textit{local amortization} (see Sec.\ \ref{sec:coverage}), the trained network has learned not only the marginal posteriors for our observed data, but also the marginal posteriors for \emph{any} data drawn from the (truncated) prior. This enables us to efficiently inspect the statistical rigor of our results, and assess fundamental statistical properties like expected coverage~\cite{hermans2021averting}. These sorts of tests are not feasible for results generated via sampling-based methods, since for each new observation a new chain is needed.
\end{enumerate}

In this paper we use \TMNRE\ to accelerate cosmological inference using simulators for the CMB power spectra. As first glance this might appear a somewhat peculiar application -- in this case the likelihood is known, and obtaining convergent \MCMC\ results using a $\Lambda$CDM cosmology is not computationally difficult. We have chosen to nevertheless apply \TMNRE\ in this context for three important reasons: $(i)$ the CMB serves as the gold standard for cosmological inference, and novel statistical tools need to be proven as robust in this context before being applied elsewhere, $(ii)$ \SBI\  trivially includes explicit likelihoods,  strong advantages even in these cases, and $(iii)$ because obtaining convergent \MCMC\ results in {\emph{extended}} cosmological scenarios {\emph{can}} be computationally prohibitive. This latter point is particularly important, as it tends to cause cosmologists to introduce unnecessary simplifications in their cosmological simulators (\eg\ by assuming massless neutrinos, simplified phase space distributions, or approximate collision terms in the Boltzmann equations). The current precision of the CMB is already sufficient in many cases to be sensitive to such approximation schemes, and thus \TMNRE\ offers a practical tool for cosmologists to robustly assess the validity of their favorite models without compromise.

The structure of this paper is as follows. In Sec.\ \ref{sec:SBITheory} we briefly review \SBI\, focusing specifically on \TMNRE\ and comparing to the alternative deep learning-based approach of \delfi\ \cite{Alsing:2018eau,Alsing:2019dvb}. In Sec.\ \ref{sec:examples} we apply this framework to different cosmological examples, using the open-source implementation of \TMNRE\ via \swyft.\footnote{ \href{https://github.com/undark-lab/swyft}{https://github.com/undark-lab/swyft}}
Each example demonstrates a unique attractive feature of our approach (or occassionally \SBI\ in general). In Sec.\ \ref{sec:cltt-warmup} we explain at length how CMB likelihoods may be reformulated as stochastic simulators (defining a ``CMB forecasting simulator''), showing that \TMNRE\ requires orders of magnitude fewer simulations than \MCMC\ to infer $\Lambda$CDM parameter constraints. In Sec.\ \ref{sec:CMB+BAO} we show how reusing simulations allows us to perform inferences for new combinations of experiments with \emph{zero} additional calls to the simulator. In Sec.\ \ref{sec:hillipop} we infer cosmological constraints for a simulator corresponding to the Planck \hillipop\ likelihood, demonstrating that the inclusion of 13 nuisance parameters does not increase the simulator cost of \TMNRE, while the cost of \MCMC\ increases by several orders of magnitude. In Sec.\ \ref{sec:global} we show that the performance of \TMNRE\ is not impeded when posteriors are significantly non-Gaussian, demonstraing on neutrino mass upper bounds. In Sec.\ \ref{sec:amort} we show how \emph{local amortization} allows us to inspect the statistical consistency of our results. Finally, in Sec.\ \ref{sec:wideprior} we demonstrate that the performance gain demonstrated by \TMNRE\ is robust even when adopting excessively wide priors.
We conclude in Sec. \ref{sec:conclusion}.

%%%%%%%%%%%%%%%%%%%%%%%%%%%%%%%%%%%%%%%%%%%%%%%%%%%%%%%
%%%%% Simulation-based inference
%%%%%%%%%%%%%%%%%%%%%%%%%%%%%%%%%%%%%%%%%%%%%%%%%%%%%%%

\section{Simulation-based inference methodology}
\label{sec:SBITheory}
The primary concern of this paper is to determine a probability distribution over model parameters $\btheta$ given an observation $\bx$. This probability distribution, called the posterior, is given by Bayes' rule as
\begin{equation}
    \label{eqn:bayes-rule}
    p(\btheta \mid \bx) = \frac{p(\bx \mid \btheta)}{p(\bx)} p(\btheta ).
\end{equation}
The likelihood of the data $\bx$ given parameters $\btheta$ is denoted $p(\bx \mid \btheta)$, $p(\btheta)$ are our prior beliefs about the parameters, and $p(\bx)$ represents the evidence of the data.

The posterior is extremely challenging to compute for most interesting physical systems, making approximation techniques commonplace. 
Traditional approaches are distinguished by whether point-wise evaluation of the likelihood function is needed. 
When the likelihood can be computed, methods such as
Markov-chain Monte Carlo (\MCMC) \cite{metropolis1953equation, hastings, green1995reversible} and nested sampling  \cite{skilling2004nested, skilling2006nested, Feroz_2009, Handley_2015, higson2019dynamic}
produce samples from the posterior distribution. Nested sampling can even estimate the evidence, which can be a valuable ingredient in model comparison. 
Alternatively, when the likelihood function is intractable one can use Approximate Bayesian computation (\ABC) \cite{sisson2018handbook, rubin1984, diggle1984monte, tavare1997}, which produces samples from an approximate posterior. However, on a practical level \ABC\ requires introducing summary statistics, a choice which can strongly influence the approximation quality and the computational cost.

Beyond these traditional approaches, recent years have seen significant progress in techniques that avoid explicitly calculating the likelihood function -- these methods fall under the umbrella of what is known as Simulation-Based Inference (\SBI) \cite{Cranmer2020}. This taxonomy includes \ABC, along with \eg\
various deep learning algorithms that train neural networks
to approximate the posterior. 
The fundamental input in \SBI\ is a stochastic simulator that maps from model parameters $\btheta$ to data $\bx$; notice that this mapping is equivalent to sampling from the distribution $\bx \sim p(\bx \mid \btheta)$ (\ie\ the likelihood).
When likelihood evaluation is not tractable, we call the distribution an \emph{implicit likelihood} function.
Using this stochastic simulator, \SBI\ methods generate a set of $N$ sample-parameter pairs $\{ (\bx^{(1)}, \btheta^{(1)}), (\bx^{(2)}, \btheta^{(2)}), \ldots, (\bx^{(N)}, \btheta^{(N)}) \}$, which can then be used as training data for a neural network that estimates either the posterior, the likelihood, or the likelihood-to-evidence ratio. We follow \cite{sbibm} and call these methods Neural Posterior Estimation (\NPE) \cite{epsilon_free, lueckmann2017flexible, greenberg2019automatic, Durkan2020,Jeffrey:2020itg}, Neural Likelihood Estimation (\NLE) \cite{papamakarios2019sequential}, and Neural Ratio Estimation (\NRE) \cite{Hermans2019, Durkan2020}, respectively. Both \NPE\ and \NLE\ estimate a normalized probability density, and are therefore restricted to architectures like mixture density networks \cite{bishop} or normalizing flows \cite{papamakarios2017masked, papamakarios2019normalizing}.
On the other hand, as we describe in Sec.\ \ref{sec:TMNRE}, \NRE\ requires only a parameterizable classifier, and is therefore highly flexible in its network architecture.
\NPE, \NLE, and \NRE\ all
produce an \emph{amortized estimate} of the posterior, which means that
once the relevant network is trained, it can sample from the posterior $p(\btheta \mid \bx)$ for any data $\bx$ without retraining.
Used appropriately, \SBI\ enables a significant reduction in computational cost when compared with sampling-based methods like \MCMC. These properties have led to the application of \SBI\ in various astrophysical contexts \cite{Alsing:2019xrx, Alsing:2018eau, dax2021real, delaunoy2020lightning}.

In paper we apply a new technique in \SBI\ called \emph{Truncated Marginal Neural Ratio Estimation} (\TMNRE) \cite{Miller:2020hua,Miller:2021hys}. The rest of this section is organized as follows. In Sec.\ \ref{sec:TMNRE}, we outline \TMNRE, describing how it extends and improves upon the standard \NRE\ approach. In Sec.\ \ref{sec:marginalization} we describe how \TMNRE's automatic marginalization leads to a large gain in simulation efficiency. In Sec.\ \ref{sec:coverage} we define coverage tests that can quickly assess the statistical consistency of results generated by \TMNRE. In Sec.\ \ref{sec:simReuse} we explain how \TMNRE\ allows the re-use of simulations across inferences.
We review related applications of \SBI\ to cosmology problems (comparing in particular to \pydelfi\ \cite{Alsing:2018eau,Alsing:2019xrx}) in Sec.\ \ref{sec:pydelfi}.

\subsection{Truncated Marginal Neural Ratio Estimation}\label{sec:TMNRE}

Here we explain the \TMNRE\ algorithm. We begin by introducing \NRE\ and theory behind likelihood-to-evidence ratio estimation. 
We then describe how to extend \NRE\ by directly targeting marginal posteriors and estimating a truncated prior over several rounds.

%%%%%%%%%%%%%%%%%%%%%%%%%%%%%%%%%%%%%%%%%%%%%%%%%%%%%%%
%%%%% Neural Ratio Estimation
%%%%%%%%%%%%%%%%%%%%%%%%%%%%%%%%%%%%%%%%%%%%%%%%%%%%%%%

\paragraph{Neural Ratio Estimation}

The target of \NRE\ is the likelihood-to-evidence ratio $r(\bx,\btheta) \equiv \frac{p(\bx \mid \btheta)}{p(\bx)} = \frac{p(\btheta \mid \bx)}{p(\btheta)} = \frac{p(\bx, \btheta)}{p(\bx) p(\btheta)}$. Note that this is equal to the ratio of the probability densities of jointly drawn sample-parameter pairs, $\bx, \btheta \sim p(\bx, \btheta)$, to marginally drawn pairs, $\bx, \btheta \sim p(\bx)p(\btheta)$. 
We introduce a binary random variable $y$ corresponding to whether a pair was drawn jointly or marginally, and let each outcome of $y$ be equally probable (in other words, we enforce that the population sizes are equal). To avoid notation clash, we use $\tilde{p}(\bx,\btheta,y)$ to denote the joint distribution including $y$ and associated quantities. Given $y$, we have
\begin{equation}
\label{eqn:classifier-conditional}
    \tilde{p}(\bx, \btheta \mid y) =
    \begin{cases}
        p(\bx, \btheta) &\text{if} \, y = 1 \\
        p(\bx) p(\btheta) &\text{if} \, y=0.
    \end{cases}
\end{equation}
Then $r(\bx,\btheta)$ may be expressed using a classifier $\tilde{p}(y=1 \mid \bx, \btheta)$, which distinguishes sample-parameter pairs drawn jointly from those drawn marginally:
\begin{equation}
\begin{aligned}
    \frac{p(\bx, \btheta)}{p(\bx) p(\btheta)} 
    = \frac{\tilde{p}(\bx, \btheta \mid y=1)}{\tilde{p}(\bx, \btheta \mid y=0)}
    &= \frac{\tilde{p}(\bx, \btheta, y=1)}{\tilde{p}(\bx, \btheta, y=0)} \\
    &= \frac{\tilde{p}(y=1 \mid \bx, \btheta)}{\tilde{p}(y=0 \mid \bx, \btheta)} \\
    &= \frac{\tilde{p}(y=1 \mid \bx, \btheta)}{1 - \tilde{p}(y=1 \mid \bx, \btheta)}.
\end{aligned}
\end{equation}
This is the so-called Likelihood Ratio Trick \cite{cranmer2015approximating}, which allows one to estimate $\frac{p(\bx~|\btheta)}{p(\btheta)}$ by estimating $\tilde{p}(y=1\mid \bx,\btheta)$.
Rewriting this expression, we have
\begin{equation}\label{eq:log_class}
    \tilde{p}(y=1 \mid \bx, \btheta) = \sigma \left( \ln r(\bx ,\btheta) \right) \, ,
\end{equation}
where we have introduced the logistic sigmoid function $\sigma(a) \equiv \frac{1}{1 + \exp(-a)}$. We train a neural network parameterized by weights $\phi$ to estimate $\tilde{p}(y=1\mid\bx, \btheta)$, in other words
\begin{equation}
    \hat{\tilde{p}}_\phi(y=1\mid \bx,\btheta)=\sigma\left(\ln \rhat_\phi(\bx,\btheta)\right).
\end{equation}
For numerical stability we do not directly train $\rhat_\phi$ but rather its logarithm, extracting $\rhat_\phi$ at the end when performing Bayesian inference.

{The estimator $\hat{\tilde{p}}_\phi(y\mid \bx,\btheta)$ is trained by maximizing its value on the observed data.
Maximizing this joint probability (over multiple data points) is equivalent to maximizing the sum of log probabilities.
In other words, we seek to minimize
\begin{equation}
    \mathbb{E}_{\tilde{p}(y, \bx, \btheta)}\left[ -\ln\hat{\tilde{p}}_\phi(y \mid \bx, \btheta) \right]
    = - \mathbb{E}_{p(\bx, \btheta)}\left[ \ln\hat{\tilde{p}}_\phi(y=1 \mid \bx, \btheta) \right] -
    \mathbb{E}_{p(\bx)p(\btheta)}\left[ \ln\left(1 - \hat{\tilde{p}}_\phi(y=1 \mid \bx, \btheta)\right) \right],
\end{equation}
}which, for our problem, means finding the parameters $\phi$ that satisfy $\argmin_\phi \ell(\rhat_\phi)$, where
\begin{equation}
\label{eqn:loss-function}
\ell(\rhat_\phi) = -\int d\bx\, d\btheta\, \left\{
        p(\bx, \btheta) \ln \sigma \left( 
            \ln \rhat_\phi(\bx \mid \btheta) 
        \right) +
        p(\bx)p(\btheta) \ln \left( 
            1 - \sigma \left( \ln \rhat_\phi(\bx \mid \btheta) \right) 
        \right)
    \right\}.
\end{equation}
This final expression is also known as the binary cross-entropy \cite{murphy, bishop}.
In practice, one draws samples from the joint distribution and trains the classifier using a sample-based approximation to the above loss function. This optimization is standard practice in deep learning and accomplished by so-called \emph{stochastic gradient descent}. Once the algorithm converges, the parameterized ratio estimator $\rhat_\phi$ allows for fast point-wise evaluation of the approximate likelihood-to-evidence ratio, enabling rejection sampling and \MCMC\ sampling from the approximate posterior.

\paragraph{Data compression}

\begin{figure}[t]
    \centering
    \includegraphics[width=0.65\linewidth]{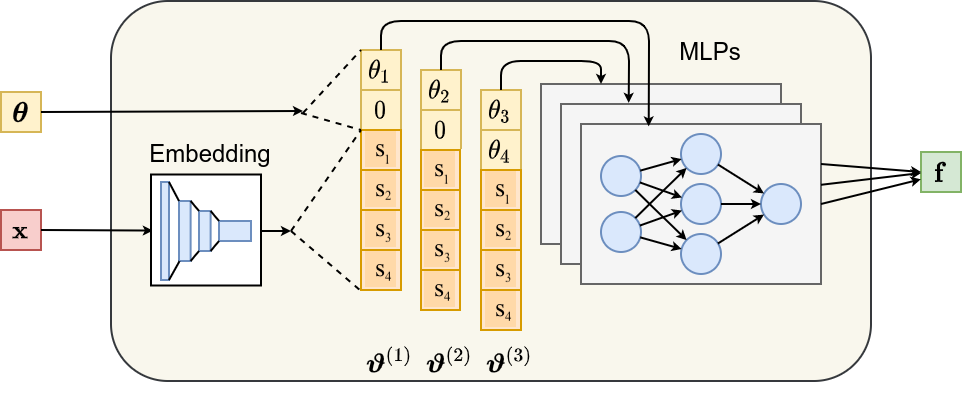}
    \caption{Illustration of the network architecture, including the ``embedding'' network $C_\phi(\bx)$ and the discrimination network (``MLP'').  Input data $\bx$ and parameters $\btheta$ are internally mapped to the marginal parameter combinations of interest, for which individual ratio estimators with a shared compression network are trained. The outputs are estimated ratios for the marginal posteriors of interest.}
    \label{fig:network}
\end{figure}

In practice, we find it beneficial to split the network structure of the ratio estimator $\hat r_\phi(\bx, \btheta)$ into two distinct components, as
\begin{equation}
   \label{eqn:rhat}
    \hat r_\phi(\bx, \btheta) = D_\phi(\bs = C_\phi(\bx), \btheta)\;.
\end{equation}
The first component is a data compression network, $C_\phi(\bx)$, which learns to compress the potentially high-dimensional data into a low-dimensional feature or summary statistics $\bs$. The structure of this network is heavily dictated by the specific properties of the data. For example, for image data, convolutional neural networks are often appropriate.  The second component, $D_\phi(\bs, \btheta)$ performs the actual ratio estimation, learning to discriminate between marginally and jointly drawn feature vectors $\bs$ and model parameters $\btheta$.  This binary classification network is typically a simple dense network with a few layers. 
In this work, we use a multi-layer perceptron (MLP) with three hidden layers, each of width 256, see Fig. \ref{fig:network} for an illustration.

Both networks are trained simultaneously, minimizing the loss function Eq.~\eqref{eqn:loss-function} as described above.  \emph{Importantly, this means that both inference and data compression happens concurrently, in contrast to other algorithms in the literature} (see discussion in Sec.~\ref{sec:pydelfi}).

It is of interest to examine how optimization selects features $\bs$ if the compression network has limited bandwidth.  Assuming that for a given compression network $C_\phi(\bx)$ the ratio estimator $r_\phi$ is fully converged, in the sense that it equals to very good approximation the posterior-to-prior ratio $r_\phi(\bx, \btheta) \simeq p(\btheta|\bs = C_\phi(\bx))/p(\btheta)$. Note that the posterior is here conditioned on the summary statistics $\bs$, since it is this summary statistics that appears in the classifier in Eq.~\eqref{eqn:rhat}.  The loss function Eq.~\eqref{eqn:loss-function} can then be written as\footnote{
   This follows directly from the definition of the JSD. In fact,
   \begin{multline}
   -2\mathbb{E}_{p(\bx)}\left[D_{JS}\left(p\left(\btheta|\bs(\bx)\right) \parallel  p(\btheta)\right)\right] \\
   =-\mathbb{E}_{p(\bx)}\left[
      D_{KL}\left(p\left(\btheta|\bs(\bx)\right) \parallel \frac12\left(p\left(\btheta|\bs(\bx)\right) + p(\btheta)\right)\right)
      +D_{KL}\left(p\left(\btheta\right) \parallel \frac12\left(p\left(\btheta|\bs(\bx)\right) + p(\btheta)\right)\right)\right]\\
   = -\int d\bx\, d\btheta\,
   \underbrace{p(\bx)p(\btheta|\bs(\bx))}_{\to p(\bx, \btheta)} \ln \frac{p\left(\btheta|\bs(\bx)\right) }{ \frac12\left(p\left(\btheta|\bs(\bx)\right) + p(\btheta)\right)}
   +p(\bx)p(\btheta) \ln \frac{p\left(\btheta\right) }{ \frac12\left(p\left(\btheta|\bs(\bx)\right) + p(\btheta)\right)}\\
   = \ell[r_\phi] - 2\ln 2 \;.
\end{multline}
The substitution in the third line is valid because the log depends on data $\bx$ only through the summary $\bs(\bx) \equiv C_\phi(\bx)$.}
\begin{equation}
\ell[r_\phi] = 2\ln 2
-2 \mathbb{E}_{p(\bx)}\left[
D_{JS}\left(p\left(\btheta|\bs = C_\phi(\bx)\right) \parallel  p(\btheta)\right)
\right]\;.
\end{equation}
Up to a constant, this is two times the negative data-averaged Jensen-Shannon divergence (JSD) between the posterior (given compressed data $\bs$) and prior.  The JSD is given by
\begin{equation}
    D_{JS}(P||Q) = \frac{1}{2}D_{KL}(P \parallel M)+\frac{1}{2}D_{KL}(Q \parallel M)\;,
\end{equation}
where $M=(P+Q)/2$, where $D_{KL}$ denotes the Kullback-Leibler divergence
\begin{equation}
    D_{KL}(P||Q) = \int dx\,P(x) \log\left(\frac{P(x)}{Q(x)}\right)\;.
\end{equation}
In other words, data summaries $\bs$ are learned such that they maximize the difference between posteriors and priors in terms of the JSD.  Note that by defnition the JSD is bound $0\leq D_\text{JS} \leq \ln 2$.

%%%%%%%%%%%%%%%%%%%%%%%%%%%%%%%%%%%%%%%%%%%%%%%%%%%%%%%
%%%%% Marginalization
%%%%%%%%%%%%%%%%%%%%%%%%%%%%%%%%%%%%%%%%%%%%%%%%%%%%%%%

\paragraph{Marginalization}
Scientific insight is often based on a low-dimensional marginalization of the posterior with nuisance parameters removed. 
\NRE\ estimates the full joint posterior $p(\btheta \mid \bx)$ over all parameters, but the marginal posterior $p(\bvartheta \mid \bx)$ is sufficient for our purposes. Here we use $\bvartheta$ to denote the parameters of interest, which are in general a low-dimensional subset of the full set of parameters $\bvartheta\subset \btheta=(\bvartheta,\boldeta)$.
We propose estimating $r(\bvartheta \mid \bx) \equiv \frac{p(\bx \mid \bvartheta)}{p(\bx)}$ directly by Marginal Neural Ratio Estimation (\MNRE) \cite{Miller:2021hys, Hermans2019}\footnote{Directly targeting marginal posteriors in the context of density estimation appears in \cite{Alsing:2019xrx,Jeffrey:2020itg}. Another technique which aims to learn all marginals simultaneously, using likelihood-to-evidence ratios, appears in \cite{rozet2021arbitrary}.}, a straightforward modification of \NRE.
The simulation procedure for \MNRE\ is exactly the same as it is for \NRE. In \MNRE\ we train $\rhat_\phi(\bx \mid \bvartheta)$, instead of the full likelihood-to-evidence ratio, by optimizing \eqref{eqn:loss-function} using a classifier restricted to only the parameters of interest $\bvartheta$. 
Evaluating the marginal posterior $p(\bvartheta \mid \bx)$ requires access to the marginalized prior $p(\bvartheta)$, but this is easy to sample and evaluate. 

From a practical perspective, a given inference involves training several classifiers (corresponding to marginal ratio estimators) in parallel. For complex data, it is generally useful to compress the data before it is input into the inference network. In this paper, we use a \emph{shared} compression network as input to \emph{all} of the classification networks. The compression and inference are performed simultaneously, as described above.  The structure is illustrated in Fig.~\ref{fig:network}.

%%%%%%%%%%%%%%%%%%%%%%%%%%%%%%%%%%%%%%%%%%%%%%%%%%%%%%%
%%%%% Truncation
%%%%%%%%%%%%%%%%%%%%%%%%%%%%%%%%%%%%%%%%%%%%%%%%%%%%%%%

\paragraph{Truncation}
Given observed data $\bx_o$, the corresponding posterior may have (significantly) narrower approximate support (in the sense that $p(\btheta\mid \bx_o)>\tilde{\epsilon}$ for small $\tilde{\epsilon}$, which we denote $\supalt{p(\btheta\mid\bx_o)}$) than the prior.
Therefore, although in \NRE\ we draw samples from the full joint distribution $p(\bx,\btheta)=p(\bx\mid\btheta)p(\btheta)$, we can learn the same ratio estimator in a local region of parameter space by drawing instead from a truncated joint distribution $p(\bx\mid\btheta)p_\Gamma(\btheta)$, where the support of $p_\Gamma(\btheta)$ is smaller than the support of $p(\btheta)$. Symbolically, given $\bx_o$ we denote the plausible subset of parameters lying above the posterior isocontour $\Tilde{\epsilon}$ as
\begin{equation}
\label{eqn:GammaDef}
    \Gamma \equiv \{ \btheta \in \supp p(\btheta) \mid p(\btheta \mid \bx_o) > \Tilde{\epsilon} \},
\end{equation}
where $\supp$ returns the support of its argument. We then avoid simulation of parameters outside $\Gamma$ by replacing our prior $p(\btheta)$ with a truncated prior 
\begin{equation}
\label{eqn:gamma}
    p_\Gamma(\btheta) \equiv  V^{-1} \indicator_{\Gamma}(\btheta) p(\btheta)\;,
\end{equation}
where $\indicator_{\Gamma}(\btheta)$ is an indicator function that is unity on $\Gamma \subset \supp p(\btheta)$ and zero otherwise, and $V^{-1}$ is a normalizing constant (which can be interpreted as the fractional volume of the truncated prior).

\begin{figure}
\centering
\includegraphics[width=\linewidth]{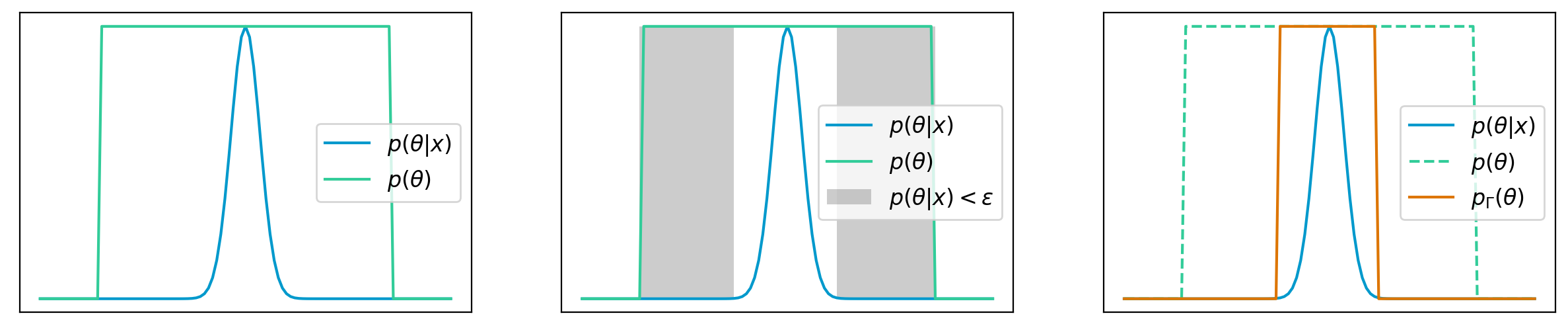}
\caption{An illustration of truncation in one dimension. Between rounds, the prior is truncated to the region where the posterior estimator exceeds a predetermined threshold.}
\label{fig:truncCartoon}
\end{figure}

Using a truncated prior $p_\Gamma(\btheta)$ focuses the simulation of parameters that feasibly lie within the highest density regions of the posterior. Truncated Marginal Neural Ratio Estimation (\TMNRE) \cite{Miller:2020hua,Miller:2021hys} implements this truncation in the context of \MNRE. In particular, we perform a round of \MNRE, giving a set of estimated marginal posteriors $\hat{p}(\bvartheta\mid \bx_o)$. These estimators are then used in an expression analogous to eqn.\ (\ref{eqn:GammaDef}) to approximate $\Gamma$. An illustration of this process is shown in Fig.\ \ref{fig:truncCartoon}.

To be more precise, we must discuss $\Gamma$.
How we choose to approximate $\Gamma$ is informed by how we suspect the approximate support of the posterior factorizes.
Note that since we will use \MNRE\ to determine our approximation, we must necessarily deal with a product of low-dimensional projections of $\Gamma$, with each projection determined by a marginal posterior for $\vartheta$.
In the simplest case, we might have an approximation that factorizes over each parameter individually, in other words $\Gamma_\pi=\supalt{p(\vartheta_1\mid \bx_o)}\times \dots \times \supalt{p(\vartheta_D\mid \bx_o)}$ for a $D$-dimensional parameter space and $\vartheta_i=\theta_i$. Here we have introduced $\pi$ to denote a collection of projections we use to describe the approximated relevant region $\Gamma_\pi$, \ie\ $\pi=\{p(\vartheta_1\mid \bx_o), \dots, p(\vartheta_D\mid \bx_o)\}$. Note that $\Gamma_\pi \supset \Gamma$. Alternatively, it may be that significant correlations are expected in the 2-dimensional marginal posterior for \eg\ $\theta_1$ and $\theta_2$. In that case a more efficient approximation is given by $\Gamma_\pi=\supalt{p\left(\vartheta_1=(\theta_1,\theta_2)\mid \bx_0\right)}\times \dots  \times \supalt{p(\vartheta_{D-1}\mid \bx)}$. For notational convenience we assume the former in the rest of this section.

We estimate $\indicator_{\GammaPi}$ by a sequence of nested indicator functions 
$\indicator_{\GammaPi^{(m)}}$ whose regions have the property $\supp{p(\theta_1)} \times \dots \times \supp{p(\theta_D)} \equiv \GammaPi^{(1)} \supset \dots \supset \GammaPi^{(M)} \supset \GammaPi$. 
The sequence iteratively approximates the indicator function $\indicator_{\GammaPi}$ in $M$ rounds and is generated via the following steps: 
\begin{itemize}
    \item Initialize $\GammaPi^{(1)} = \supp{p(\theta_1)} \times \dots \times \supp{p(\theta_D)}$, \ie\ we start with the unconstrained prior.
    \item In each round $1 \leq m \leq M$, we train $D$, one dimensional ratio estimators ${\hat r_{d, \GammaPi^{(m)}}(\bx \mid \theta_{d})}$ using data from within the constrained region, $\btheta \in \GammaPi^{(m)}$.
    The estimated marginal posterior is $\hat p_{\GammaPi^{(m)}} (\theta_{d} \mid \bx) = 
    \hat r_{d, \GammaPi^{(m)}}(\bx \mid \theta_{d}) p_{\GammaPi^{(m)}} (\theta_{d})$.
    To this end, we perform \MNRE, setting $\bvartheta = \theta_{d}, \; d \in \{1, 2, \ldots, D\}$ using the constrained prior $p_{\GammaPi^{(m)}}(\btheta)$ with $N^{(m)}$ training samples per round.
    \item For each round $m<M$, we estimate the indicator function for the next round using the approximated posteriors, the analogy of eqn.\ (\ref{eqn:GammaDef}) being
    \begin{equation}
        \GammaPi^{(m+1)} = 
        \left\{ \btheta \in \GammaPi^{(m)}  
    	\; \middle| \;
    	\forall d: \frac{\hat p_{\GammaPi^{(m)}}(\theta_d \mid \bx_o)}{\max_{\theta_d} 
    	\hat p_{\GammaPi^{(m)}}(\theta_d \mid \bx_o)} > \epsilon \right\}\;.
    \end{equation}
    
    \item The last round is determined either when $m=M$ or when a stopping criterion is reached.
    The stopping criterion is defined by the ratio of consecutive truncated prior masses. It is satisfied when the sequence of truncated priors have the property $$\int \indicator_{\GammaPi^{(m)}}(\btheta) p(\btheta) d\btheta / \int \indicator_{\GammaPi^{(m-1)}}(\btheta) p(\btheta) d\btheta > \beta.$$ We often set $\beta = 0.8$.
    
    \item Using the data from this final constrained region, we can approximate any marginal posterior of interest $p(\bvartheta \mid \bx)$. We emphasize that the data already generated during the truncation phase can be reused to learn arbitrary marginals of interest. When higher accuracy likelihood-to-evidence ratio estimates are needed, the user can simulate from the truncated region.
\end{itemize}

For considerations of the geometry of truncation in more than 1 dimension, see \cite{code}. In most of the examples in Sec.\ \ref{sec:examples} we perform \MNRE\ (\ie\ a single round) rather than \TMNRE. We demonstrate \TMNRE\ (\ie\ multiple rounds) in Sec.\ \ref{sec:wideprior}. An illustration of the truncation in that example is shown in Fig.\ \ref{fig:widePriors}.

With the details of \TMNRE\ in hand, we now address how our algorithm fares in comparison to alternative sampling-based inference methods, specifically highlighting the benefits offered by \TMNRE\ in terms of efficiency, scalability, and testability.

\subsection{``Cutting to the chase" by estimating marginal posteriors}
\label{sec:marginalization}

\begin{figure}[h]
    \centering
    \includegraphics[width=0.7\linewidth]{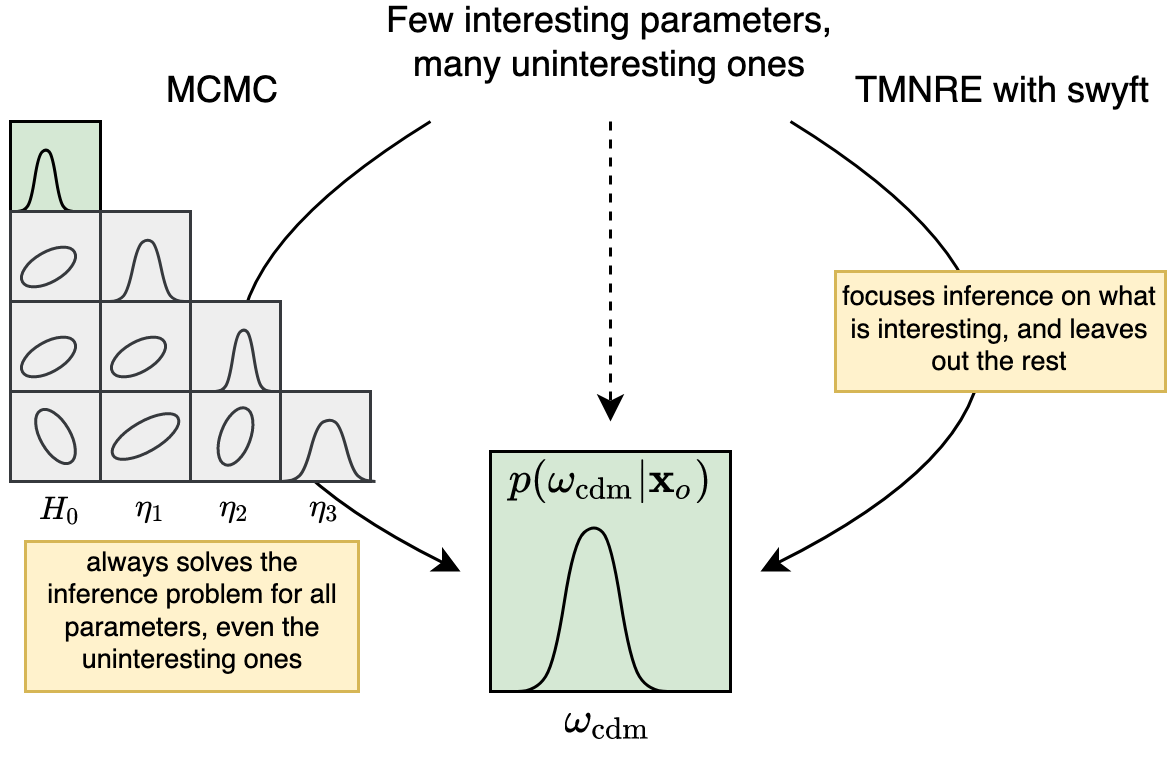}
    \caption{
    \TMNRE\ can handle complex models and directly estimate any marginal posterior of interest, which typically requires far fewer simulation runs than estimating the full joint posterior.
    }
    \label{fig:efficiency}
\end{figure}

As we have mentioned, scientific conclusions about the measured value of a parameter of interest, say the cold dark matter energy density $\omega_{cdm}$, are based on \textit{marginal} posteriors $p(\omega_{cdm} \mid \bx_o)$.  As illustrated in Fig.~\ref{fig:efficiency}, \SBI\ methods such as \TMNRE\ allow us to directly estimate marginal posteriors of interest, whereas likelihood-based methods must always first estimate the joint posterior for all model parameters. More specifically, the marginal posterior is related to the joint posterior, $p(\omega_{cdm}, \eta_1, \dots, \eta_N | \bx_o)$, via integration,
\begin{equation}
    p(\omega_{cdm}\mid\bx) =  \int d\boldsymbol{\eta} \,
    p(\omega_{cdm}, \eta_1, \dots, \eta_N \mid \bx)\;.
\end{equation}
Likelihood-based inference techniques do not directly generate samples from the one-dimensional or two-dimensional posteriors of interest, but instead always generate first samples from the full high-dimensional joint posterior.  Those can then be marginalized (\ie\ projected onto $\omega_{cdm}$ direction) to generate samples from the marginal posterior $p(\omega_{cdm}\mid \bx)$.

Estimating low-dimensional marginals of interest with simulation-based techniques can therefore be significantly more computationally efficient than estimating the joint posterior with likelihood-based methods, especially when the number of parameters and/or modes is high. In fact, the numbers of samples required for convergence of likelihood-based methods increases very quickly (typically cubically \cite{Handley_2015} or even exponentially~\cite{Feroz_2009}, in the case of nested sampling) as the number of model parameters grows.\footnote{For a pedagogical introduction to Markov Chain Monte Carlos see Ref.~\cite{2019arXiv190912313S}.}  In contrast, the required number of simulations for simulation-based techniques can be nearly independent of the number of parameters involved if only low-dimensional marginals are of interest.

As a result, likelihood-free inference methods, like \TMNRE, enable the estimation of any marginal posteriors of interest with orders of magnitude fewer simulations runs than likelihood-based methods. We demonstrate this in the context of CMB data throughout Sec.\ \ref{sec:examples}.

{Note that we are not claiming that the end goal of cosmological data analysis is to produce 6 one-dimensional posteriors corresponding to each $\Lambda$CDM parameter. Indeed, it is often of interest to assess higher-dimensional structure. For example, if several experiments give high-dimensional posteriors that are in tension with each other, this can be obscured by a low-dimensional projection.\footnote{If one is interested in quantifying a tension between two datasets, this is in fact possible in the framework of neural ratio estimation without examining marginal posteriors. Instead, one can consider the contrastive problem of distinguishing between $(x_1, x_2)\sim p(x_1, x_2)$ and $(x_1, x_2) \sim p(x_1)p(x_2)$ (where $x_i$ denotes the data from a particular experiment), which proceeds in the full (high-dimensional) parameter space. We leave this avenue for future work.} Additionally, it is generally of interest to assess degeneracies between constraints on different parameters (hence the typical ``triangle plot'').  Finally, although we are using a language in which it is useful to ``ignore'' nuisance parameters (i.e. learn a low-dimensional distribution), note that in general we propose learning low-dimensional posteriors for the nuisance parameters and using them in the truncation process. Therefore at the end posteriors for nuisance parameters may be inspected to assess consistency and extract information about e.g.\ foregrounds, systematics, and astrophysics. Indeed, from the perspective of assessing the $S_8$ tension by examining the $(\sigma_8, \Omega_m)$ the other $\Lambda$CDM parameters are ignored, and these are clearly relevant to cosmology!}

\subsection{``Getting it right'' through coverage tests for simulation-based inference}\label{sec:coverage}

\begin{figure}[h]
    \centering
    \includegraphics[width=0.7\linewidth]{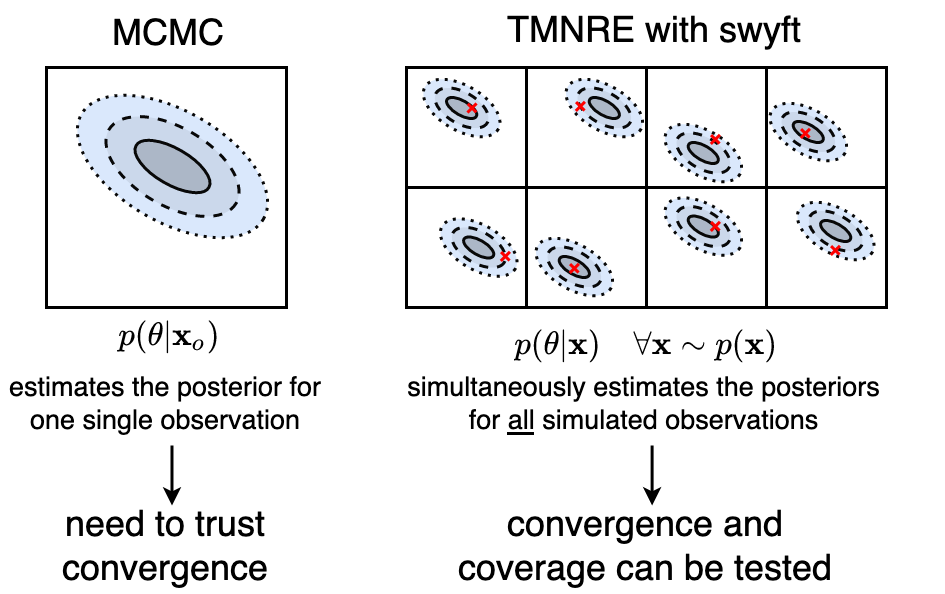}
    \caption{
    Testability: Many \SBI\ methods, including \TMNRE\, do not estimate just a single posterior, but \emph{all} of them simultaneously (``amortization''). This enables the user to test the reliability of the inference results.
}
    \label{fig:testability}
\end{figure}

Although the internal workings of the trained network are usually hard to understand without detailed inspection, the enormous evaluation speed of trained neural networks enables important consistency tests that we introduce below. These tests are generally infeasible for likelihood-based methods, where the estimate of one single posterior is already very costly (illustrated in the left panel of Fig.~\ref{fig:testability}). For that reason, the expected statistical properties of Bayesian inference results (like biases, credibility) are usually not directly tested but are instead inferred based on convergence criteria~\cite{roy2019convergence, betancourt2018convergence} and asymptotic properties of stationary Markov Chains.  
On the other hand, trained inference networks (like the ratio estimators in \TMNRE), constitute essentially ultra-fast black-box data analysis pipelines.  They take as input any observed data $\bx\sim p(\bx)$, and very quickly generate the corresponding posteriors $p(\btheta\mid\bx)$.  This is illustrated in the right panel of Fig.~\ref{fig:testability}.

Let us denote with $\Theta_{p(\bvartheta\mid\bx)}(1-\alpha)$ the $1-\alpha$ highest posterior density region. For example, a $95\%$ highest posterior density region would have an expected error rate of $\alpha = 0.05$. In other words, when randomly drawing observations $\bx, \bvartheta \sim p(\bx\mid\bvartheta)p(\bvartheta)$, we expect the true parameters $\bvartheta$ to fall outside of the 95\% region in $5\%$ of the cases.
Following Ref.~\cite{hermans2021averting}, the expected coverage probability of the $1-\alpha$ highest posterior density region (HPDR) of some estimated posterior $\hat p(\bvartheta|\bx)$ is given by
\begin{equation}
\label{eqn:highest-posterior-density-region}
    1-\hat\alpha = \mathbb{E}_{p(\boldsymbol{\vartheta}, \boldsymbol{x})}\left[\mathds{1}\left[\boldsymbol{\vartheta} \in \Theta_{\hat{p}(\boldsymbol{\vartheta} \mid \boldsymbol{x})}(1-\alpha)\right]\right]
\end{equation}
This quantity can be interpreted as both expected Bayesian credibility as well as expected Frequentist coverage probability of $\hat p(\bvartheta|\bx)$~\cite{hermans2021averting}.  It enables us to estimate the \emph{actual} error rate $\hat \alpha$ of a $1-\alpha$ highest posterior density region of some estimated posterior $\hat p(\bvartheta|\bx)$. 

\paragraph{Coverage figure construction.}
In the case where $\hat p(\bvartheta|\bx) = p(\bvartheta|\bx)$, it follows that $\hat \alpha(\alpha) = \alpha$. We can hence use $\hat\alpha$ as function of $\alpha$ as quality measure for posteriors estimated with \TMNRE. Since small values of $\alpha$ are of particular interest, corresponding to posterior regions with high mass, we find it convenient to reparameterize $\alpha$ in terms of a new variable $z$.
We define $z$ as the $1-\frac12\alpha$ quantile of the standard normal distribution, for instance $\alpha = 0.05$ corresponds to $z = 1.96$.  The commonly quoted ``$1\sigma$'', ``$2\sigma$'' and ``$3\sigma$'' regions, correspond to $z = 1, 2, 3$ and hence have $1-\alpha = 0.6827, 0.9545, 0.9997$.

\begin{figure}[h]
    \centering
    \includegraphics[width=0.9\linewidth]{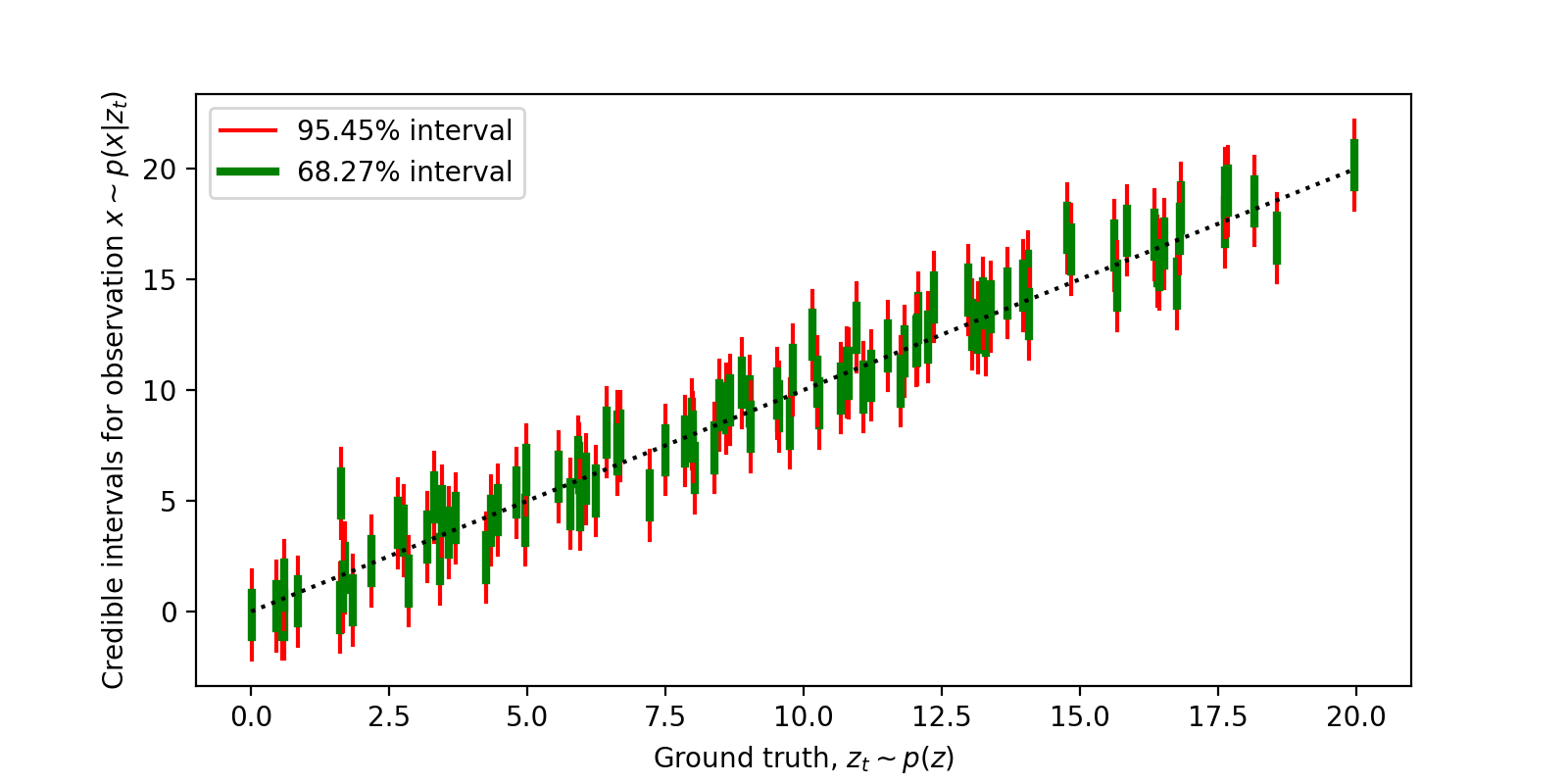}
    \caption{
       Illustration of a coverage test for a one-dimensional interval.  The ``ground truth'' parameter values $z_t$ for a range of test simulations are drawn from the prior, $z_t \sim p(z)$.  The credible intervals are derived from the corresponding simultaneously generated mock observations, $x\sim p(x|z_t)$.  One then obtains the number of times the ground truth is contained in, say, the 68.27\% interval (66 times in our 100 examples), or the 95.45\% interval (95 times).
    }
    \label{fig:coverage_illustration}
\end{figure}

We estimate empirically the actual error rate $\hat\alpha$ (and the corresponding $\hat z$) for nominal $1-\alpha$ regions as follows (see Fig.~\ref{fig:coverage_illustration} for an illustrative example).  We generate $n$ samples from the joint model $\bx, \bvartheta \sim p(\bx, \bvartheta)$, and count how often a $1-\alpha$ highest posterior density region predicted by the trained network does \emph{not} contain the true parameter $\bvartheta$. We denote this number by $k$, which follows a binomial distribution with fractional probability $\hat \alpha$ and total number of draws $n$.  The mean is then simply given by $\hat \alpha = k/n$. We propose to estimate uncertainties of this quantity using the so-called Jeffreys interval, which is effectively the  Bayesian credible interval obtained from a non-informative Jeffreys prior for a binomial distribution~\cite{JeffreysInterval}.\footnote{
More specifically, the interval is obtained from the $68.27\%$ central interval of a Beta distribution with parameters $(n-k+\frac12, k+\frac12)$.
One of the advantages of the Jeffreys interval is that it is approximately equally tailed, with intervals evenly distributed above and below the true value. This convenient property is preserved even when reparametrizing the inferred parameter in terms of $\hat z$.  
}

\begin{figure}[h]
    \centering
    \includegraphics[width=0.49\linewidth]{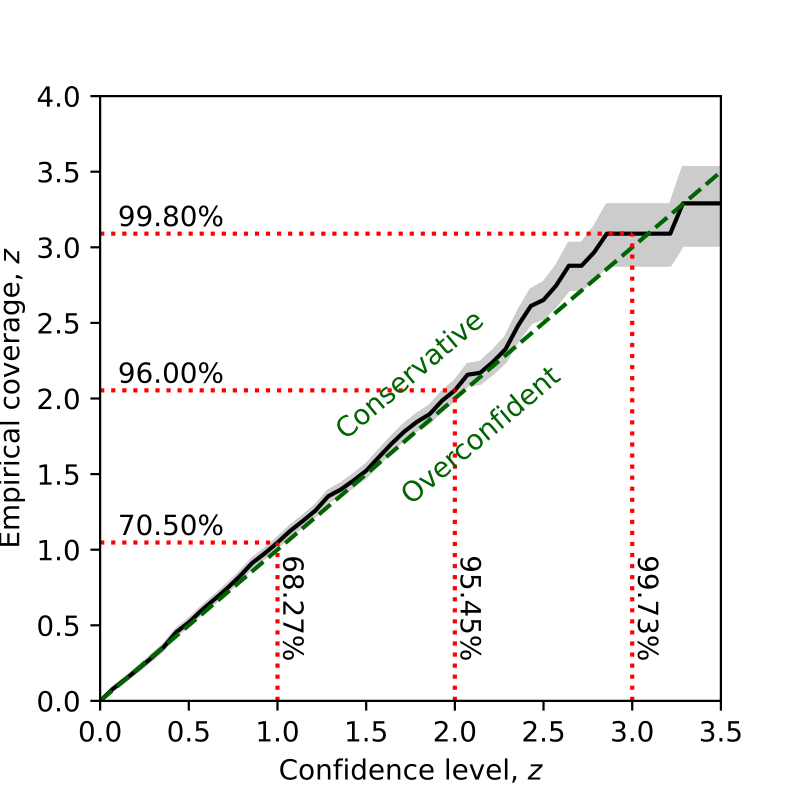}
    \includegraphics[width=0.49\linewidth]{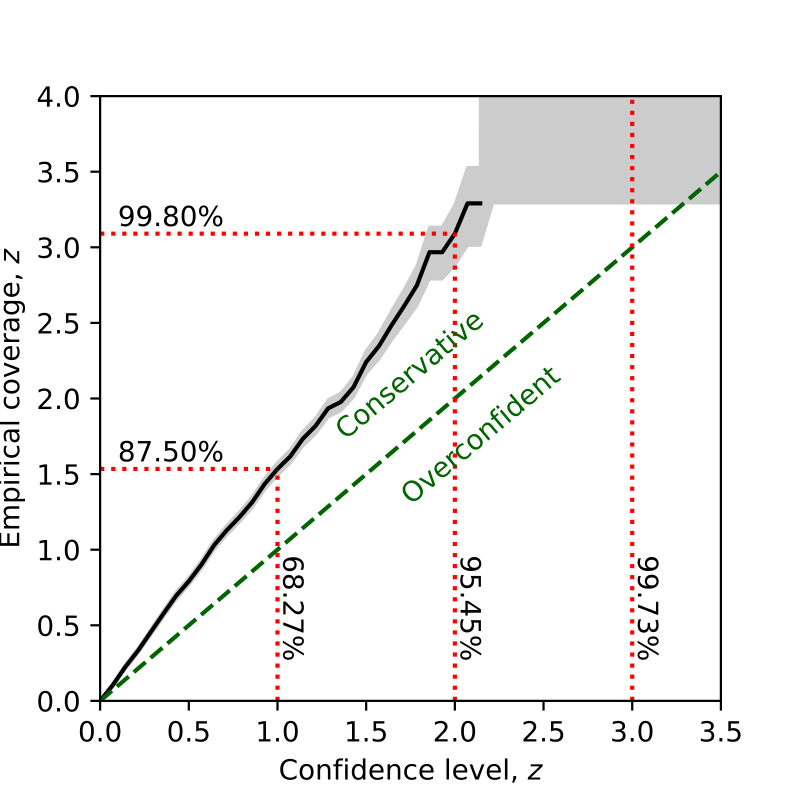}
    \caption{
    Empirical coverage as function of the confidence level, as function of the significance level $z$ (see text).  In cases where $\hat z > z$, estimated credible contours are conservative and contain the true value with a frequency higher than nominally expected.  The nominal probabilities $1-\alpha$ are shown in the figure as vertical text, the empirical estimates $1-\hat\alpha$ are shown as horizontal text.  
    \textit{Left panel:} Converged network with good coverage. \textit{Right panel:} Conservative coverage, as typically shown by networks during training before convergence.
    }
    \label{fig:coverage_converged}
\end{figure}

We summarize the results of such a test in Fig.~\ref{fig:coverage_converged}.
The empirical coverage $\hat z$ is shown as function of the confidence levels $z$ for 1000 test simulations.  We also show the error band based on Jeffreys interval, with the widths determined by the number of test simulations used.  We furthermore indicate the empirical coverage in terms of $1-\hat \alpha$ (horizontal numbers) for typical values of the confidence level $1-\alpha$ (vertical numbers).  In this particular example, the $68.27\%$ ($95.45\%$) highest posterior density region contains the ground truth value in $70.50\%$ ($96.00\%$) of the cases, and is hence slightly conservative, or equivalently the confidence intervals slightly over-cover (and are thus conservative).
In the case of perfect coverage, one would expect the black line and the green dashed line in Fig.~\ref{fig:coverage_converged} to perfectly overlap. Should the black line lie above the green line, the confidence intervals over-cover (and are thus conservative; see the right panel of Fig.~\ref{fig:coverage_converged}), while if the black line lies below the confidence intervals under-cover (\textit{i.e.}~are over-confident).  

Figures like Fig.~\ref{fig:coverage_converged} can be used to test and validate the correct statistical behavior of an inference network, which is particularly useful when one does not have a ground-truth against which to compare the results (\eg\ in the examples given below, our ground truth is taken to be a converged \MCMC).  For over-conservative or overconfident networks, those figures can be also used to provide corrections to quoted credible levels.
In Sec.\ \ref{sec:amort}, we use empirical coverage plots in order to double-check and confirm the convergence of our posterior estimators for CMB data. { Note that similar tests and plots have appear in the literature in \cite{dax2021real,hermans2021averting,Jeffrey:2021fcg}. Additionally, while what we propose is a necessary condition for the calibration of our approximate posteriors, it is not sufficient. Moreover, we only test the consistency of our approximation of the posterior given the simulation model and prior choice. Developing more tests in order to trust results generated by \SBI\ is worthwhile and ongoing work.}

\subsection{Zero-cost inference through simulation re-use} \label{sec:simReuse}

    Simulation runs in likelihood-based methods are commonly difficult to re-use, because the generation of proposal points for simulations is directly intertwined with the inference process for a specific observation.  In contrast, the generation of training data through simulation runs and the training of inference networks are distinct steps in the 
    context of \TMNRE.\footnote{{Note that this is also a feature of other variational \SBI\ methods.}}
    This makes the re-use of previous simulations almost trivial: the same training data, sampled from the generative model, can be used train multiple networks, it can be combined with other training data, or it can sub-sampled to change priors or reduce training data.

In the context of \TMNRE, and for the purpose of this paper, we will adopt a simple sample cache model (for a more the general case, see Ref.~\cite{Miller:2020hua}).  Let us assume we generated training $\mathcal{S}_1$ data consisting of $N$ samples from the truncated prior $p_{\Gamma_1}(\bvartheta)$.  Now we are interested in generating $M$ samples from the truncated prior $p_{\Gamma_2}(\bvartheta)$, where we assume that $\Gamma_2 \subset \Gamma_1$.  We can now simply re-use all samples in $\mathcal{S}_1$ that are inside $\Gamma_2$ (let us assume there are $N'\leq N$ of those), and then add to those $M-N'$ samples from $p_{\Gamma_2}(\bvartheta)$.  In cases where the truncation region does not change, we can re-use all previous samples and hence simulations.

In Sec.\ \ref{sec:CMB+BAO} we re-use simulations to efficiently compare distinct experimental configurations.
We demonstrate in Sec.\ \ref{sec:wideprior} that this scheme allows us to quickly zoom from wide initial priors into the parameter region relevant for a given observation, while keeping the number of required simulations low.
\subsection{Comparison with other methods}
\label{sec:pydelfi}
{ Although our focus in this work is a deep learning-based method for \SBI, progress continues to be made in existing approaches as well. For example, \ABC\ continues to be successfully applied to astrophysics, with public codes incorporating importance sampling via e.g.\ Population Monte Carlo \cite{Akeret:2015uha,Ishida:2015wla} or Sequential Monte Carlo \cite{jennings2017astroabc}. Similarly, approaches that leverage Gaussian approximations of likelihoods have been used and combined with Gaussian processes \cite{Mootoovaloo:2020ott,Heavens:2020spq}. Gaussian process regression has also been used to enable Bayesian optimization \cite{Leclercq:2018who,Rogers:2018smb,Takhtaganov:2019ywj}. Along the lines of Gaussian approximations, including perturbations around Gaussian likelihoods has also been pursued \cite{Sellentin:2014zta,Leclercq:2019aud}.}

On the deep learning side, applications to cosmology have appeared on several fronts \cite{Zhao:2021ddh,Makinen:2021nly,Villaescusa-Navarro:2021pkb,Villaescusa-Navarro:2021cni}. Most applications have employed density estimation methods (\delfi) \cite{Alsing:2018eau,Alsing:2019xrx}. In the rest of this section, we review \delfi, specifying to its implementation via \pydelfi, and describe how it relates to \TMNRE.

\delfi\ constitutes Neural Likelihood Estimation (\NLE) or Neural Posterior Estimation (\NPE). For notational simplicity, we refer to \delfi\ in \NLE\ mode. \NLE\ learns the sampling distribution $\bx \sim p(\bx |\btheta)$ of data $\bx$ as a function of model parameters $\btheta$. 
\NLE\ becomes increasingly difficult when the dimension of the data is large, or the data distribution is highly complex \cite{papamakarios2019normalizing}.
This problem can be alleviated by compressing the data to some simple low-dimensional summary statistics before performing inference with \NLE. 
In \pydelfi\ this is often done by introducing a compression network such as an Information Maximizing Neural Network, which maximizes the Fisher information of the compression variables \cite{Charnock_2018}.  
This compression network is typically trained upfront before inference, generally at additional simulation costs.
The network architecture used for the subsequent \NLE\ inference step is restricted to density estimation networks such as mixture density networks \cite{bishop} or normalizing flows \cite{papamakarios2017masked, papamakarios2019normalizing}
(while simpler network structures such as multi-layer perceptron (MLP) are not an option).
With compressed data summaries in hand, \pydelfi\ uses one of two approaches to draw new training data relevant for a given observation (sometimes coined ``active learning''). In the first approach, \pydelfi\ uses Sequential \NLE\ (\SNLE) \cite{papamakarios2019sequential}, in which the parameters for each batch of simulations are drawn from a proposal density based on the current posterior estimator, and the neural density estimators are re-trained after each simulation batch. Commonly the proposal density is the current posterior estimator itself. In the second approach, each new batch of simulations is given by a deterministic acquisition function that seeks to maximize both relevance (\ie\ large posterior density) and uncertainty (usually quantified by comparing the estimates of multiple networks).

\pydelfi\ and \TMNRE\ are similar in several respects:
	\begin{itemize}
		\item \emph{Amortization}: in non-active learning mode (\ie\ \NLE\ or \MNRE), networks are manifestly amortized throughout the full prior. In active learning mode, \TMNRE\ is manifestly amortized throughout the truncated prior, and the network trained by \eg\ \SNLE\ can still be evaluated throughout the prior. 
		Both approaches learn quantities that are independent of the proposal distribution.
		While the proposal distribution means that \SNLE\ is expected to be less accurate away from the observation of interest, the empirical tests afforded by amortization may still be carried out.
		{This amortization is a unique aspect of methods that train variational approximations of Bayesian quantities that can be rapidly evaluated or sampled.}
		
		\item \emph{Sampling}: after training, both approaches use e.g.\ \MCMC\ to sample from the (marginal) neural posterior. This evaluation is generally very fast thanks to the massive evaluation speed and parallelization afforded by neural networks, see \eg\ \cite{SpurioMancini:2021ppk}. This neural posterior sampling can also be combined with sampling from additional explicit likelihoods as needed.
		
		\item \emph{Compression}: for complex data, it is advantageous for both methods to employ a compression network. The training of the compression network is slightly different in the two cases, see below.
		
		\item  \emph{Marginalization}: both approaches can directly target marginal posteriors \cite{Alsing:2019dvb,Jeffrey:2020itg,Miller:2020hua,Miller:2021hys}.  In the context of \pydelfi, this marginalization can be integrated into the compression step by constructing ``nuisance-hardened'' summary statistics \cite{Alsing:2019dvb}. \pydelfi\ has been shown to outperform \MCMC\ even when targeting the full joint posterior \cite{Alsing:2018eau}. {The ability to directly target marginal posteriors is only possible for variational methods.}
		
		\item \emph{Active learning}: both approaches preferentially simulate with model parameters belonging to some region of interest. In \TMNRE\ this is done via truncation. In \pydelfi\ this is done either via \SNLE\ or optimization of a deterministic acquisition function. { As mentioned previously, Bayesian optimization is also relevant to non-neural \SBI\ \cite{Leclercq:2018who,Rogers:2018smb,Takhtaganov:2019ywj}.}
	\end{itemize}
	
	We therefore emphasize that many of the attractive features of \TMNRE\ we demonstrate via examples in Sec.\ \ref{sec:examples} also have analogues using \pydelfi\ or other variational \SBI\ approaches.
	%Both approaches are generally superior to \MCMC.
	
    There are two main differences between \TMNRE\ and \pydelfi:
	\begin{itemize}
		\item \emph{Ratio vs.\ density estimation}: ratios (as learned by \TMNRE) can be estimated with simple battle-tested classification networks, which are often easier to train than density estimation models (as used by \pydelfi) \cite{papamakarios2019normalizing}.  However, especially in the case where the functional shape of posteriors are sufficiently simple and Gaussian, the inductive bias of density estimation models can increase the simulation efficiency of density-estimation based approaches~\cite{Miller:2021hys}. Moreover, density estimation learns a cheap and reusable generative model for the sampling distribution of the data, which is not true for \NRE\ parameterized via \eg\ an MLP.  One option, which we leave to future work, is to parameterize ratio estimation using a flow-based model, in which case these benefits are shared.
		
		The difference between density and ratio estimation affects how the compression network is trained.
		In \pydelfi, the compression network is trained beforethe inference network, while in \TMNRE\ the compressed representation is learned simultaneously with the inference network.

		\item \emph{Interplay between marginalization and active learning dynamics:} in \TMNRE, active learning is accomplished through truncation, while in \pydelfi, it is accomplished via e.g.\ sequential methods. These approaches are especially different when targeting marginal posteriors. Note first that when a joint likelihood-to-evidence ratio or likelihood is estimated, the proposal distribution for model parameters $\btheta$ drops out.\footnote{For \NRE, this is only approximately true, since the evidence $p(\bx)=\int d\btheta~ p(\bx\mid\btheta)p(\btheta)$ depends on $p(\btheta)$; however, this is a constant for fixed $\bx$.} On the other hand, when targeting marginals so that $\btheta=(\vartheta,\eta)$, the proposal distribution for $\eta$ is important. For example, the marginal likelihood is defined by $p(\bx\mid\vartheta)=\int d\eta~p(\bx\mid \vartheta, \eta)p(\eta)$. The nuisance parameters $\eta$ must be sampled according to the correct distribution. In truncation, the marginal posteriors of the nuisance parameters are learned and used to define a truncated prior \emph{including the nuisance parameters}. This strategy is successful even when the priors for nuisance parameters are very wide, see \cite{Coogan:2020yux} for an example.
		On the other hand, when \pydelfi\ uses sequential methods for marginals, the nuisance parameters are sampled from the full prior distribution. In practice the user aims to reduce the dependence of the summary statistics on the nuisance parameters via nuisance-hardening. In the optimally nuisance-hardened case there is no dependence of the summary statistics on the nuisance parameters, but this may be difficult to achieve in extreme cases.
	\end{itemize}
		
		More research is needed to compare in detail the performance of \NRE, \NLE, and \NPE\ approaches in various settings. We expect that further progress can be made by combining insights from approaches that are regarded as distinct within the current taxonomy.

\section{Cosmological applications}\label{sec:examples}

 In this section we apply \TMNRE\ to several cosmological examples. Each example is designed to showcase a particular attractive feature of the algorithm:

\begin{enumerate}
    \item  \textbf{Directly estimating marginals is much more efficient than estimating the joint and then marginalizing.} In Sec.\ \ref{sec:cltt-warmup} we descibe at length how to define a simulator for CMB forecasting, and show that by learning marginal rather than joint posteriors, \TMNRE\ converges much more quickly than \MCMC.
    \item \textbf{Rerunning the analysis with different experimental configurations does not require new simulations.} In Sec.\ \ref{sec:CMB+BAO} we demonstrate simulation reuse in the context of comparing CMB inferences with and without BAO measurements.
    \item \textbf{Increasing the number of nuisance parameters does not increase the number of required simulations.} In Sec.\ \ref{sec:hillipop} we perform inference for a realistic Planck simulator, showing that the 13 varying nuisance parameters do not impact the simulator cost for \TMNRE.
    \item \textbf{Non-Gaussian posteriors are a non-issue.} In Sec.\ \ref{sec:global} we show that \TMNRE\ accurately and cheaply infers an upper bound for the sum of neutrino masses, the marginal posterior of which is highly non-Gaussian.
    \item \textbf{Amortization allows cheap consistency checks.} In Sec.\ \ref{sec:amort} we demonstrate the rigorous statistical testing of \TMNRE\ results, which can be performed even when the ground truth posteriors are unknown.
    \item \textbf{Wide priors are not a problem.} In Sec.\ \ref{sec:wideprior} we take a wide prior and show that \TMNRE's truncation efficiently identifies the relevant region of parameter space, so that simulations are not wasted on irrelevant parameters.
\end{enumerate}

\subsection{A CMB forecasting simulator}
\label{sec:cltt-warmup}
To use \SBI, we require a simulator. In this section we define very explicitly a simulator with the same statistical content as CMB likelihoods commonly used for forecasting. We couple this simulator to \TMNRE\ to show that:

\begin{center}
\textbf{Directly estimating marginal posteriors is much more efficient than estimating the joint posterior and then marginalizing.}
\end{center}

We now describe the simulator at some length. Implementations of the corresponding likelihood are available via \texttt{MontePython}, in the \texttt{Likelihood\_mock\_cmb} class \cite{Brinckmann:2018owf}. Our notation follows the discussion in \cite{Perotto:2006rj}. In particular, the multipole coefficients $a_{\ell m}$ of CMB maps receive contributions from the CMB signal $s_{\ell m}$ and experimental noise $n_{\ell m}$
\begin{equation}
    a_{\ell m}^P=s_{\ell m}^P+n_{\ell m}^P
\end{equation}
where the index $P$ runs over temperature, polarization, and weak lensing degrees of freedom. For simplicity, we restrict to $P=T,E$ in our analysis, but the extension is straightforward. In the full sky limit, the two-point correlation of $a_{\ell m}$ is given by
\begin{equation}
    \braket{a_{\ell m}^{P*}a_{\ell' m'}^{P'}}=\left(C_{\ell}^{PP'}+N_{\ell}^{PP'}\right)\delta_{\ell \ell'}\delta_{mm'}\equiv \overline{C}_{\ell}^{PP'}\delta_{\ell \ell'}\delta_{mm'}
\end{equation}
where $\braket{\cdots}$ denotes an ensemble average. For specified cosmological parameters, the power spectra $C_{\ell}^{PP'}$ can be computed via a Boltzmann code such as \texttt{CLASS} \cite{Blas:2011rf}. The effects of spatially uniform Gaussian white noise and a Gaussian beam give rise to a noise contribution
\begin{equation}
    N_{\ell}^{PP'}\equiv \braket{n_{\ell m}^{P*}n_{\ell m}^{P'}}=\delta_{PP'}\theta_{\rm fwhm}^2 \sigma_{P}^2 \exp\left(\ell(\ell+1)+\frac{\theta_{\rm fwhm}^2}{8\ln 2}\right)
\end{equation}
where $\theta_{\rm fwhm}$ is the full width at half maximum of the beam and $\sigma_P$ is the root-mean-square of the instrumental noise. In this section we will use values for $N_{\ell}^{PP'}$ compatible with the Planck experiment (accessible within \texttt{MontePython} as \texttt{fake\_planck\_realistic}) \cite{DiValentino:2016foa}.

Given $\overline{C}_\ell(\btheta)^{PP'}$ (making explicit the dependence on underlying cosmology parameters, collectively denoted $\btheta$), the $a_{\ell m}$ are distributed according to the likelihood
\begin{equation}
    p({\bf a}\mid\btheta)\propto \frac{1}{|\overline{C}(\btheta)|^{1/2}} \exp\left(-\frac{1}{2}{\bf a}^\dag[\overline{C}(\btheta)^{-1}]{\bf a}\right)
\end{equation}
where ${\bf a}=\{a_{\ell m}^T,a_{\ell m}^E\}$ and $\overline{C}(\btheta)^{-1}$ denotes matrix inversion. In general, one may sample from this using the Cholesky decomposition of the covariance matrix. For fixed $\ell$ and $m$, one has
\begin{equation}
    \left(\begin{matrix}a_{\ell m}^T\\ a_{\ell m}^E\end{matrix}\right)=L\left(\begin{matrix}n_{1}\\
     n_{2}\end{matrix}\right)
\end{equation}
where $n_i$ are sampled from normal distributions with unit variance $n_i\sim \mathcal{N}(0,1)$ and $L$ is a matrix satisfying $LL^T=\left(\begin{matrix}\overline{C}_\ell^{TT}& \overline{C}_\ell^{TE}\\\overline{C}_\ell^{TE}& \overline{C}_\ell^{EE}\end{matrix}\right)$. For each $\ell$ there are $2\ell+1$ modes. Therefore for a single realization of the universe, $\overline{C}_\ell$ can only be determined to finite precision. Given a realization of $a_{\ell m}^P$, the maximum likelihood estimator for $\overline{C}_\ell$, which we denote $\hat{C}_\ell^{PP'}$, is given by
\begin{equation}
    \hat{C}_\ell^{PP'}=\frac{1}{2\ell+1}\sum_{m=-\ell}^{\ell}a_{\ell m}^{P*} a_{\ell m}^{P'}=\frac{1}{2\ell+1}\left(a_{\ell 0}^P a_{\ell 0}^{P'}+2\sum_{m=1}^{\ell}a_{\ell m}^{P*}a_{\ell m}^{P'}\right)
\end{equation}
Therefore the map $\btheta\to \overline{C}(\btheta)\to \hat{C}$ is stochastic in the last step. For fixed $\overline{C}_{\ell}$, the Gaussian distribution of $a_{\ell m}$ propagates to a Wishart distribution for $\hat{C}_\ell$ \cite{Percival:2006ss}. Explicitly, given $\overline{C}_\ell^{PP'}$, the likelihood for $\hat{C}(\btheta)$ is given by
\begin{equation}\label{eqn:likelihoodCl}
    -2\ln p\left(\hat{C}(\btheta)\mid\overline{C}\right)=\chi^2_{\rm eff}=\sum_{\ell}\left(2\ell+1\right)\left[\frac{D}{|\overline{C}|}+\ln\frac{|\overline{C}|}{|\hat{C}|}-2\right]
\end{equation}
where $D=\overline{C}_\ell^{TT} \hat{C}_\ell ^{EE}+\hat{C}_\ell^{TT} \overline{C}_\ell ^{EE}-2 \overline{C}_\ell^{TE}\hat{C}_\ell^{TE}$ and we have used our freedom in shifting the log-likelihood by a constant so that $\ln  p\left(\hat{C}(\btheta)=\overline{C}\mid\overline{C}\right)=0$. The Wishart distribution is significantly non-Gaussian at low $\ell$, which is necessary to reflect the fact that the autocorrelations $\hat{C}_{\ell}^{PP}$ are positive. At large $\ell$, the Wishart distribution can be approximated as a Gaussian distribution with
\begin{equation}\label{eqn:forecastCovCl}
    \textrm{Cov}_{C_\ell}=\frac{2}{2\ell+1}\left(\begin{matrix}
    \left(\overline{C}_\ell^{TT}\right)^2& \overline{C}_\ell^{TT}\overline{C}_\ell^{TE}& \left(\overline{C}_\ell^{TE}\right)^2\\
    \overline{C}_\ell^{TT}\overline{C}_\ell^{TE} & ~\frac{1}{2}\left(\overline{C}_\ell^{TT}\overline{C}_\ell^{EE}+\left(C_\ell^{TE}\right)^2\right)~& \overline{C}_\ell^{TE}\overline{C}_\ell^{EE}\\
    \left(\overline{C}_\ell^{TE}\right)^2& \overline{C}_\ell^{TE}\overline{C}_\ell^{EE}& \left(\overline{C}_\ell^{EE}\right)^2
    \end{matrix}\right)
\end{equation}
Sampling from this distribution is more efficient than sampling from the full Wishart distribution, especially at large $\ell$ where the approximation is valid. We will take the Gaussian approximation starting at $\ell=52$.

Finally, we account for partial sky coverage by modifying the effective degrees of freedom at each $\ell$. We rescale the covariance matrix (\ref{eqn:forecastCovCl}) by $2\ell+1\to f_{\rm sky}\left(2\ell+1\right)$ at high $\ell$. At low $\ell$, we round $f_{\rm sky} (2\ell+1)$ to the nearest integer and sample this many $a_{\ell m}$, subsequently using the maximum likelihood estimator to determine $\hat{C}_\ell^{PP'}$. (We show a power spectrum analysis with explicit mask in App.\ \ref{app:mask}.)

\begin{figure}
    \centering
    \includegraphics[width=\textwidth]{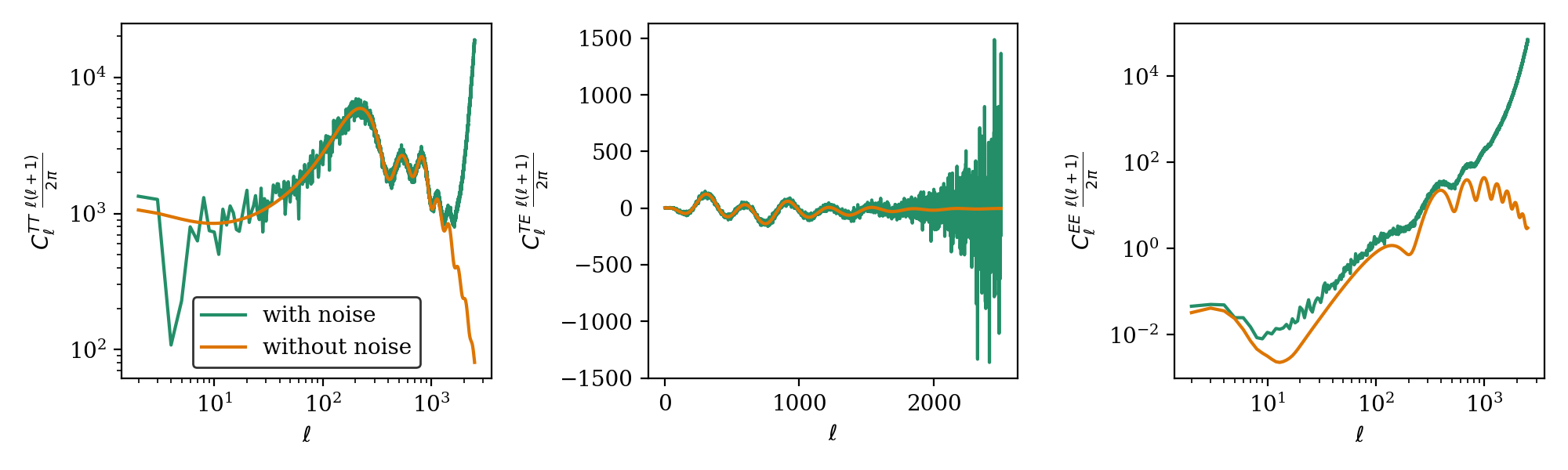}
    \caption{Samples drawn from the CMB forecasting simulator. Here ``without noise'' denotes the output of \class\ (\ie\ $C_{\ell}^{PP'}$) and ``with noise'' includes both experimental noise and cosmic variance (\ie\ $\hat{C}_\ell^{PP'}$).}
    \label{fig:samples}
\end{figure}

In other words, a simulator that is equivalent to the likelihood (\ref{eqn:likelihoodCl}) is defined by the following steps:
\begin{enumerate}
    \item Given cosmological parameters, compute $C_{\ell}^{PP'}$ via a Boltzmann code.
    \item Add experimental noise: $\overline{C}_{\ell}^{PP'}=C_\ell^{PP'}+N_{\ell}^{PP'}$
    \item From $\overline{C}_\ell^{PP'}$, compute $\hat{C}_\ell^{PP'}$ by sampling $a_{\ell m}$ and the maximum likelihood estimator at low $\ell$ (equivalently, sampling the Wishart distribution). At large $\ell$, use $\hat{C}_\ell^{PP'}=\overline{C}_\ell^{PP'}+n_\ell^{PP'}$, where $n_\ell^{PP'}$ is sampled from the multivariate normal distribution (\ref{eqn:forecastCovCl}).
\end{enumerate}
Finally, in likelihood-based approaches uncertainties are incorporated in the form of the likelihood and therefore not sampled explicitly (for forecasting, sometimes even for the fiducial observation) \cite{Perotto:2006rj}. Note that in \SBI\ we must sample from the noise model. Samples from this simulator are shown in Fig.\ \ref{fig:samples}.  We now proceed to perform inference with this simulator.

\subsubsection{Inference setup and results}

Using this simulator, we perform inference for a $\Lambda$CDM cosmology with $\ell_{\rm max}=2500$. We use noise $N_\ell^{PP'}$ given by \texttt{fake\_planck\_realistic} \cite{DiValentino:2016foa}. In the context of forecasting, we are free to fix the fiducial cosmology. We choose a fiducial cosmology\footnote{For this experiment, we use Asimov data, i.e.\ $\hat{C}_\ell^{PP'}=\overline{C}_\ell^{PP'}$, as is customary for forecasting. Since the likelihood for $\overline{C}_\ell^{PP}$ is non-Gaussian at low multipole, we rescale these modes accordingly for the fiducial data. This is mainly to show that the typical forecasting workflow is possible. In later sections the fiducial data includes non-trivial noise contributions.}
\begin{equation}
    \btheta^0=\left(\omega_{\rm b},\omega_{\rm cdm},100\theta_s,\ln\left(10^{10}A_s\right),n_s,\tau\right)=\left(0.0224,~ 0.12,~ 1.0411,~ 3.0753,~ 0.965,~ 0.054\right)
\end{equation}
A Gaussian approximation to the likelihood can be computed using the Fisher information matrix
\begin{equation}
    F_{ij}\equiv -\left.\frac{\partial^2 \ln \mathcal{L}}{\partial\theta_i\partial\theta_j}\right|_{\btheta^0}
\end{equation}
In this approximation, the formal error on the parameter $\theta_i$ is given by
\begin{equation}
    \sigma_i^F=\sqrt{(F^{-1})_{ii}}
\end{equation}
For this simulator with known likelihood, the Fisher matrix can be computed:
\begin{equation}\label{eqn:fisher}
    F_{ij}=\sum_{\ell=2}^{\ell_{\rm max}}\sum_{PP',QQ'}\frac{\partial C_\ell^{PP'}}{\partial\theta_i}\left({\rm  Cov}_{C_\ell}^{-1}\right)\frac{\partial C_{\ell}^{QQ'}}{\partial\theta_j}
\end{equation}
where ${\rm Cov}_{C_\ell}$ is given by (\ref{eqn:forecastCovCl}) (including a factor of $f_{\rm sky}$). We compute the Fisher matrix using a two-sided numerical approximation for the derivatives $\partial_{\theta_i}C_\ell^{PP'}$. We use the resulting $\sigma_i^F$ to define a uniform prior on each cosmological parameter $\theta_i$
\begin{equation}
    \theta_i\sim \mathcal{U}\left([\theta_i^0- 5\sigma^F_i,\theta_i^0 + 5\sigma^F_i]\right)
\end{equation}
The final ingredient to our setup is a compression network. Since there are only 6 varying parameters in our model, it is reasonable to expect that the data can be optimally compressed into 6 features.
%\footnote{One can also consider the various noise contributions (organized by e.g.\ a singular value decomposition of the covariance matrix) as free parameters of the model. In this case the complexity is increased, } 
When the likelihood is known, the ``score function'' $\vec{t}=\nabla_{\btheta}\ln \mathcal{L}(\hat{C}(\btheta)|\btheta)$ compresses the data into one summary per parameter, such that the Fisher information is preserved \cite{Alsing:2017var,Alsing:2018eau}. In the case of a Gaussian likelihood, score compression is equivalent to \moped\ \cite{Heavens:1999am} (if $\btheta$ determines the mean) or the optimal quadratic estimator \cite{Tegmark:1996bz} (if $\btheta$ determines the covariance). If a likelihood is not known, neural compression schemes can be trained to find reduced representations of the data that maximize information content \cite{Charnock_2018}.
 
We choose to remain agnostic about the compression, apart from noting that within our prior, the power spectra $\overline{C}_\ell^{PP'}$ are well-described by a linear approximation in the cosmological parameters.\footnote{This is not strictly necessary for using a linear compression, since the subsequent classifier can learn non-linearities of the posteriors. \TMNRE\ accurately reproduces non-Gaussian (i.e.\ non-linear) features in the posterior of this simulator.} We therefore use as a compression network a linear map from 7497 to 15 features. We use $15$ features since this allows more capacity for the network without significantly increasing computational cost.

\begin{figure}[ht]
    \centering
    \includegraphics[width=\textwidth]{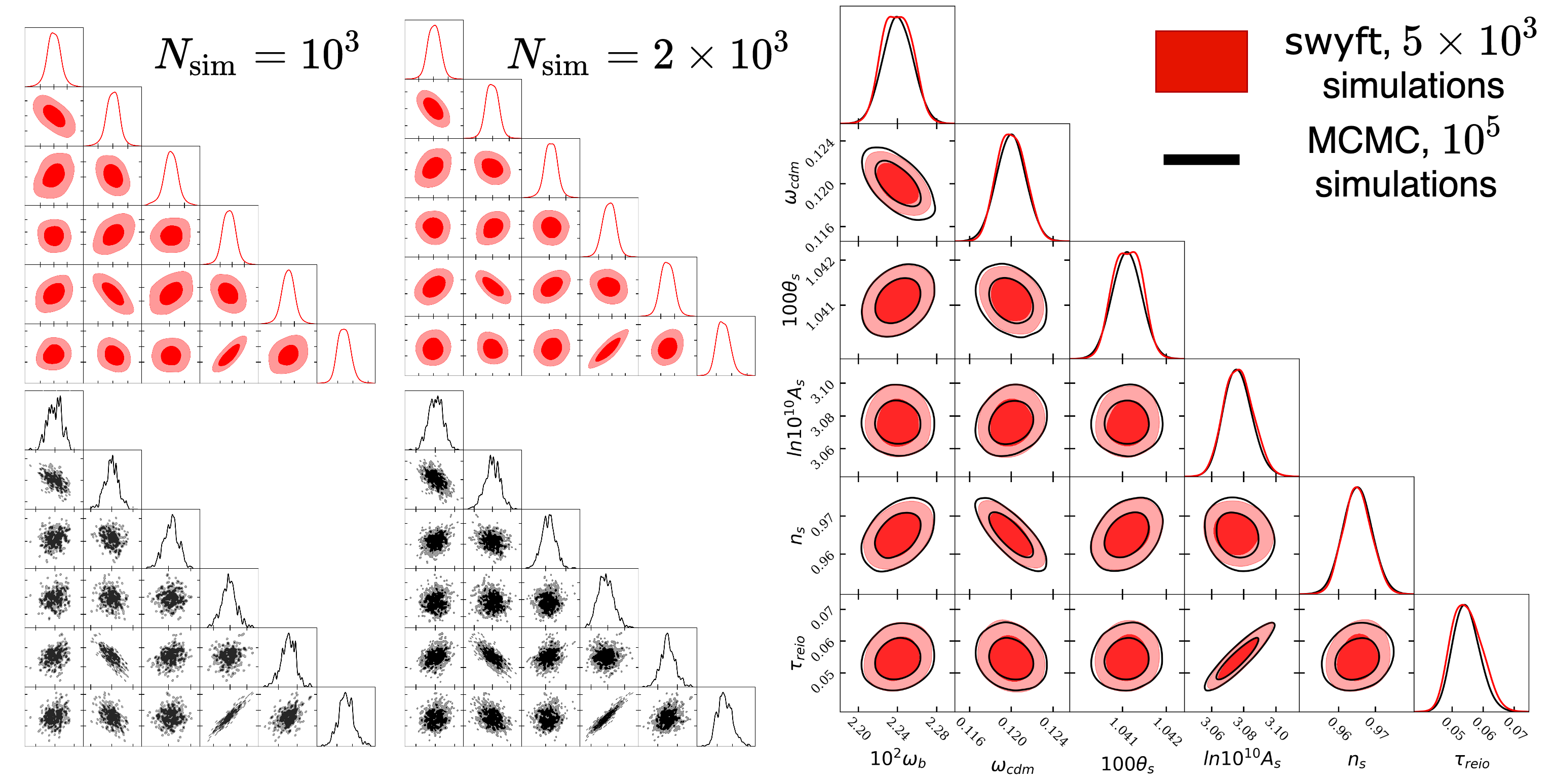}
    \caption{Left: convergence of marginal posteriors for \TMNRE\ vs.\ \MCMC\ for the CMB forecasting simulator. Since \MCMC\ must sample the full joint posterior, it is slow to converge. On the other hand, \TMNRE\ directly learns the marginal posteriors, and begins to converge quickly. Right: comparison of \TMNRE\ at $5\times 10^3$ simulation to ground-truth \MCMC\ with $10^5$ simulations. \TMNRE\ agrees with the ground-truth to excellent precision and at significantly reduced simulator cost. Note that tick mark values are the same for all plots.}
    \label{fig:CMBForecasting}
\end{figure}

We use \TMNRE\ to train \emph{every} 1- and 2-dimensional marginal posterior using $N_{\rm sim}\leq 5000$ simulations. {Training all of the relevant posteriors in parallel takes about 10 minutes on our GPU. Additionally, we note that the same 5000 simulations are used to train every posterior.}
These results are shown in Fig.\ \ref{fig:CMBForecasting}, as well as posteriors derived via \MCMC\ for $N_{\rm sim}\leq 10^5$ simulator calls. The \MCMC\ chain had an acceptance rate of $\sim 0.3$, so that $10^5$ simulator calls were required to produce $\sim 3\times 10^4$ samples. \MCMC, which \emph{samples} the \emph{joint} (6-dimensional) posterior, requires this many samples in order to produce converged $2\sigma$ contours, while \TMNRE, which \emph{learns} the \emph{marginal} posteriors, produces results in excellent agreement at significantly reduced simulator cost.\footnote{A common criterion for the convergence of \MCMC\ results is that the Gelman-Rubin statistic $\hat{R}-1$ is sufficiently small \cite{gelman1992inference}, for example $<0.01$. We employ this as a necessary condition for converged results. In addition, we require that 1- and 2-$\sigma$ credible regions are visually converged when the kernel smoothing using to produce plots is fixed. Note that this sometimes means running our chains beyond $\hat{R}-1 < 0.01$ (in particular for high-dimensional examples). We use the same kernel smoothing to produce plots for both \MCMC\ and \TMNRE; the advantage to \TMNRE\ is that once trained (on a potentially small number of simulations), a very large number of samples may be drawn from the variational posterior. \label{foot:converge}} Moreover, the number of simulations given to \TMNRE\ can be decreased without significantly impacting the quality of the posteriors; we found that at $N_{\rm sim}=3000$ simulations, \TMNRE\ produces 1-dimensional marginal posteriors whose central values and $1\sigma$ credible regions were accurate to within $0.05\sigma$ of the ground truth.

\subsection{Comparing experimental configurations}
\label{sec:CMB+BAO}
It is often of interest to study how constraints on parameters (or from a forecasting perspective, expected sensitivies) change when different combinations of experiments are considered. 
Given the large number of upcoming cosmological experiments at any given time, the number of relevant experimental configurations can be daunting (e.g.\ 30 configurations of CMB and LSS experiments for forecasting constraints on neutrino masses in \cite{Brinckmann:2018owf}). In the context of \MCMC, each experimental configuration must be treated with its own chain(s) in order to preserve Markov properties. The situation, however, is significantly improved for \TMNRE.

\begin{center}
\textbf{Re-using simulations allows for additional inferences at \emph{zero} simulator cost.}
\end{center}

This opens the opportunity to perform massive global scans with significantly reduced simulator cost. As demonstrated in the previous section, for a single inference \TMNRE\ can be orders of magnitude more simulator-efficient than \MCMC. By reusing simulations across inferences, \TMNRE\ can gain another order of magnitude of simulator-efficiency relative to sampling methods.

More explicitly, for \MCMC, the simulator cost for generating posteriors for $N_{\rm config}$ experimental configurations scales as $N_{\rm config}N_{\rm sim,\MCMC}$, where $N_{\rm sim, \MCMC}$ is the typical number of simulator calls for \MCMC\ to converge for an individual experimental configuration. In the case where simulations can be reused, the computational cost for \TMNRE\ scales as $N_{\rm sim, \TMNRE}\ll N_{\rm sim, \MCMC}$.

There are two alternative approaches for reusing simulations in the standard framework. One is the reweighting of \MCMC\ chains via an alternative likelihood. One drawback of this approach is that deviations beyond 1- or 2-$\sigma$ can not be addressed. Moreover, reweighting requires either recomputing the data (in which case one might as well run a new chain) or having these data saved. As we have seen, \MCMC\ requires orders of magnitude more simulator calls than \TMNRE, so storing these simulations quickly becomes a burden.

A second approach is to use simulations to train emulators, enabling ultra-fast likelihood evaluations \cite{Fendt:2006uh,Euclid:2020rfv,SpurioMancini:2021ppk,Mootoovaloo:2021rot,Rogers:2018smb}. {Alternatively, one may pair expensive and cheap simulations or models to construct approximate likelihoods in a simulation-efficient way \cite{Hall:2018umb,Chartier:2020pmu}.} While these approaches are also attractive, there are a few drawbacks. In particular, these approaches are generally possible only for restricted summary statistics such as power spectra. { Even when neural networks are trained to output \eg\ full boxes of galaxy surveys \cite{He:2018ggn,AlvesdeOliveira:2020yix,Kaushal:2021hqv}, these methods are typically validated on low-order or otherwise restricted information;}
%Even when emulators are designed to output \eg\ full boxes of galaxy surveys \cite{tassev2013solving,tassev2015scola,howlett2015picola}, they are validated on low-order information;
the degree to which higher-order information can be trusted (and used as input for inference) is more limited. Even in the case of a power spectrum analysis, when a new cosmological model is developed, it is more efficient to do inference with \TMNRE\ before training an entire emulator.

\begin{figure}[ht]
    \centering
    \includegraphics[width=0.75\textwidth]{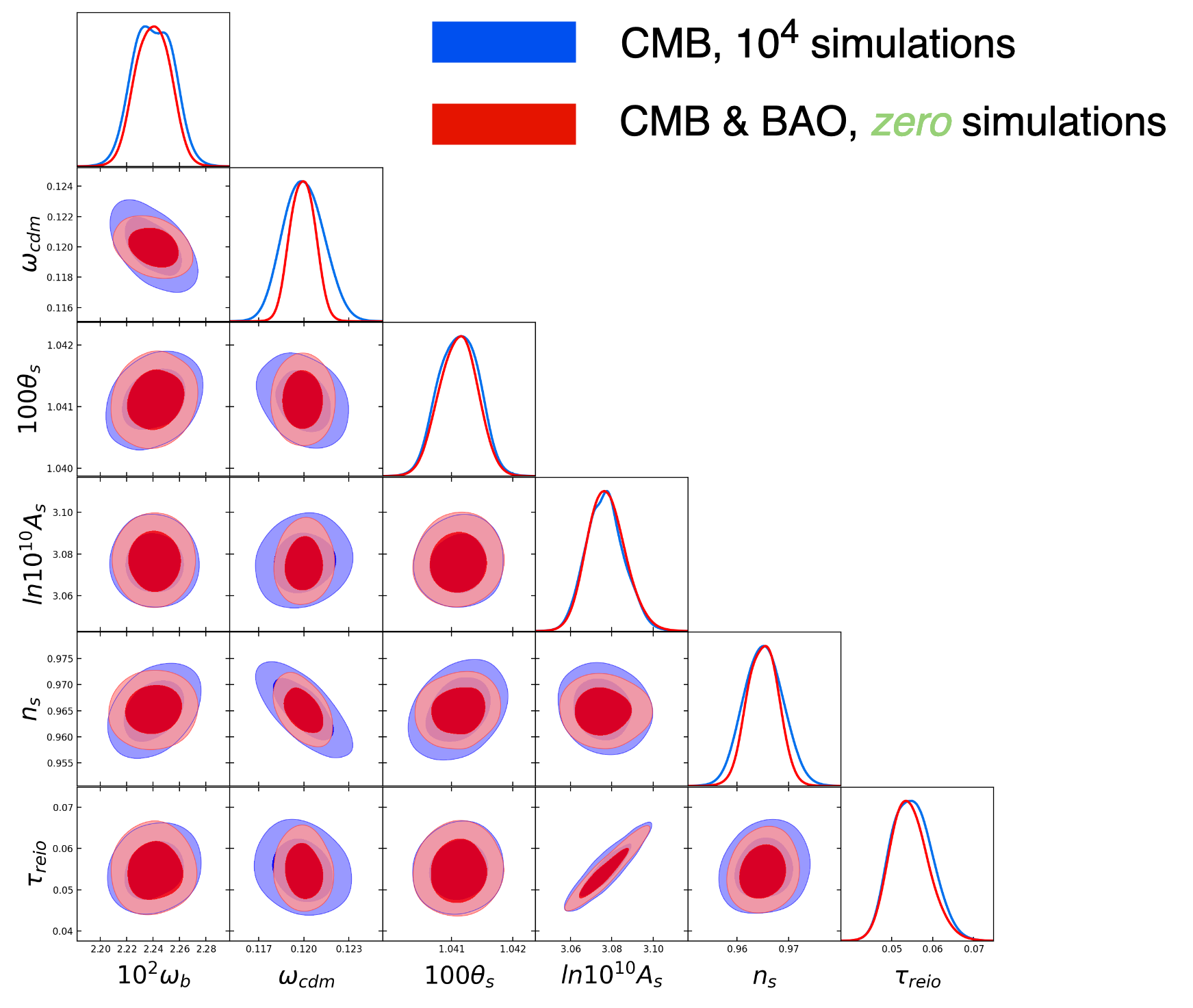}
    \caption{Demonstration of simulation reuse with \TMNRE. We infer posteriors corresponding to the forecasting CMB simulator of Sec.\ \ref{sec:cltt-warmup} with $N_{\rm sim}=10^4$ simulations. We subsequently use the \emph{same} simulations to infer posteriors for the combination of CMB and BAO data.}
    \label{fig:simReuse}
\end{figure}

We demonstrate the principle of simulation reuse across different experimental configurations in Fig.\ \ref{fig:simReuse}. We first use \TMNRE\ to compute posteriors for the CMB forecasting simulator with $N_{\rm sim}=10^4$ simulations. Then we use the \emph{same} simulation bank to train \TMNRE\ for 
an alternative experimental configuration
corresponding to the combination of CMB and BAO data. 
To generate the BAO data, when performing simulations for the first inference, we include a simulator component corresponding to BAO measurements at $z=0.106,~0.32,~0.57,~0.15$,
with sensitivities proportional to the 6dF Galaxy Survey \cite{Beutler_2011}, Sloan Digital Sky Survey (SDSS-III) Baryon Oscillation Spectroscopic Survey (BOSS) data release 11 and CMASS data \cite{Anderson_2014}, and SDSS Main Galaxy Sample \cite{Ross_2015} respectively.\footnote{We shrink the noise uncertainties by a factor of two for the sake of visualization.}\footnote{The same logic applies to considering an upgraded CMB experiment. In this case, we may simply change the simulator's noise model. Note that it is efficient to save the ``noise-free'' simulations and only add noise later, when performing the inference. Often this second step is much quicker than the initial simulation.} The simulator corresponding to these measurements takes the BAO value derived by \class\ and adds Gaussian noise inferred from experimental sensitivities.
It should be clear from the previous section that likelihood approaches to LSS that use more information from the power spectra (e.g. the full shape of the galaxy power spectrum \cite{Philcox:2020vvt,Chudaykin:2020aoj}) can also be phrased as simulators.

We perform inference using the same settings as in Sec.\ \ref{sec:cltt-warmup}, including a linear compression network that maps from 7497 (7501 for CMB+BAO) to 15 features. 
As in the previous section, for \emph{each} inference \MCMC\ requires $N_{\rm sim,\MCMC}=\mathcal{O}(10^5)$ simulator calls for converged $2\sigma$ contours, so that for $N_{\rm config}=2$ experimental configurations, the total cost is $2\times 10^5$. On the other hand, after the first inference, \TMNRE\ requires \emph{zero} new simulator calls, so that the total simulator cost saturates at the single-inference value, in this case $\mathcal{O}(10^4)$. This logic clearly extends to the case of more experimental configurations, the number of which often exceeds an order of magnitude for realistic cosmological inferences.

\subsection{Realistic Planck simulator}\label{sec:hillipop}
Classical inference becomes more difficult as the instrument model becomes more realistic. This is because realistic instrument models involve extra parameters related to the calibration of the instrument, foreground residuals, etc. These quantities, which are not interesting to the scientific problem at hand, are called \emph{nuisance parameters}. Sampling cost for the joint posterior scales poorly with the addition of extra parameters.
%However, \TMNRE\ eats nuisance parameters for breakfast.
{On the other hand, \TMNRE\ is not siginifcantly affected by nuisance parameters.}

\begin{center}
    \textbf{Directly learning marginal posteriors alleviates sampling problems associated with the curse of dimensionality.}
\end{center}

We now turn to \hillipop\footnote{\href{https://github.com/planck-npipe/hillipop}{https://github.com/planck-npipe/hillipop}} \cite{Couchot:2016vaq}, which is a multifrequency CMB likelihood for Planck data. In particular, \hillipop\ treats high-$\ell$ modes within a Gaussian approximation, including cross-correlation between Planck's 100, 143, and 217 $\textrm{Ghz}$ split-frequency maps. The model covers multipoles from $\ell=30$ to $\ell=2500$, and includes contributions from galactic dust, the cosmic infrared background, thermal and kinetic Sunyaev-Zeldovich emission, and point sources. 

Since the likelihood is Gaussian, it can be naturally rephrased in simulator language, as in Sec.\ \ref{sec:cltt-warmup}. Namely, \hillipop\ computes the likelihood using a $\chi^2$
\begin{equation}
    \chi^2 = (d-\mu)^T C^{-1}(d-\mu)
\end{equation}
where $d$ is computed by binning, adding foregrounds, etc.\ to a spectrum computed from a Boltzmann code and $\mu$ is the observed data. We therefore peel off $d$ from the likelihood and use this as the output of our simulator, drawing noise from the covariance matrix $C$ as needed.

We use the temperature autocorrelation part of the simulator, which has 10816 bins and 13 varying nuisance parameters in addition to 6 cosmological parameters. We choose this likelihood because (i) it has many nuisance parameters (ii) it has a native Python implementation.\footnote{Among other Planck likelihoods, \texttt{plik} \cite{Planck:2019nip} has a similar number of nuisance parameters but no native Python implementation. The nuisance-marginalized version of this likelihood is available in Python as \texttt{plik\_lite} via Cobaya \cite{Torrado:2020dgo} or \cite{Prince:2019hse}. We have checked that inference for the simulator corresponding to this likelihood is as roughly as easy with \TMNRE\ as the CMB forecasting simulator of Sec.\ \ref{sec:cltt-warmup}.}

{The model takes 6 cosmological parameters and 13 varying nuisance parameters. }
For priors on the cosmological parameters, we use uniform distributions with $\pm 5\sigma$ from Planck 2018 \cite{Planck:2018vyg}. Since this is a high-multipole simulator, we impose a Gaussian prior on the optical depth, $\tau=0.0544\pm 0.0073$, corresponding to Planck low-multipole constraints.\footnote{For progress towards simulation-based inference for low-$\ell$ modes, see \cite{Prince:2021fdv}.} For priors on the nuisance parameters, we use the recommendations shipped with \hillipop: Gaussian priors for all 6 varying calibration parameters (with $\sigma=0.0025$ for $A_{\rm Planck}$ and $\sigma=0.002$ for varying $c_{X}$) and Gaussian priors (with $\sigma=0.2$) for all foreground amplitudes other than $A_{\rm kSZ}$ and $A_{\rm SZ\times CIB}$, which are both uniformly drawn from the interval $[0,10]$.\footnote{See \href{https://github.com/planck-npipe/hillipop/blob/master/examples/hillipop\_example.yaml}{https://github.com/planck-npipe/hillipop/blob/master/examples/hillipop\_example.yaml} for the full list of priors.}

{In order to have control over our setup, we infer posteriors for a fiducial spectrum generated by our simulator rather than Planck data. For all parameters with Gaussian priors except $\tau$ and $A_{\rm radio}$, we displace the fiducial parameter by $-0.5\sigma$ from the mode. For parameters with uniform priors in the interval $[l, h]$, we set the parameter equal to $l+\frac{3(h-l)}{10}$. For $\tau$ and $A_{\rm radio}$ (where the difference between the prior and posterior is most dramatic in \cite{Couchot:2016vaq}), we displace the fiducial parameter from the mode of the prior by $+2\sigma$.}

\begin{figure}[ht]
    \centering
    \includegraphics[width=\textwidth]{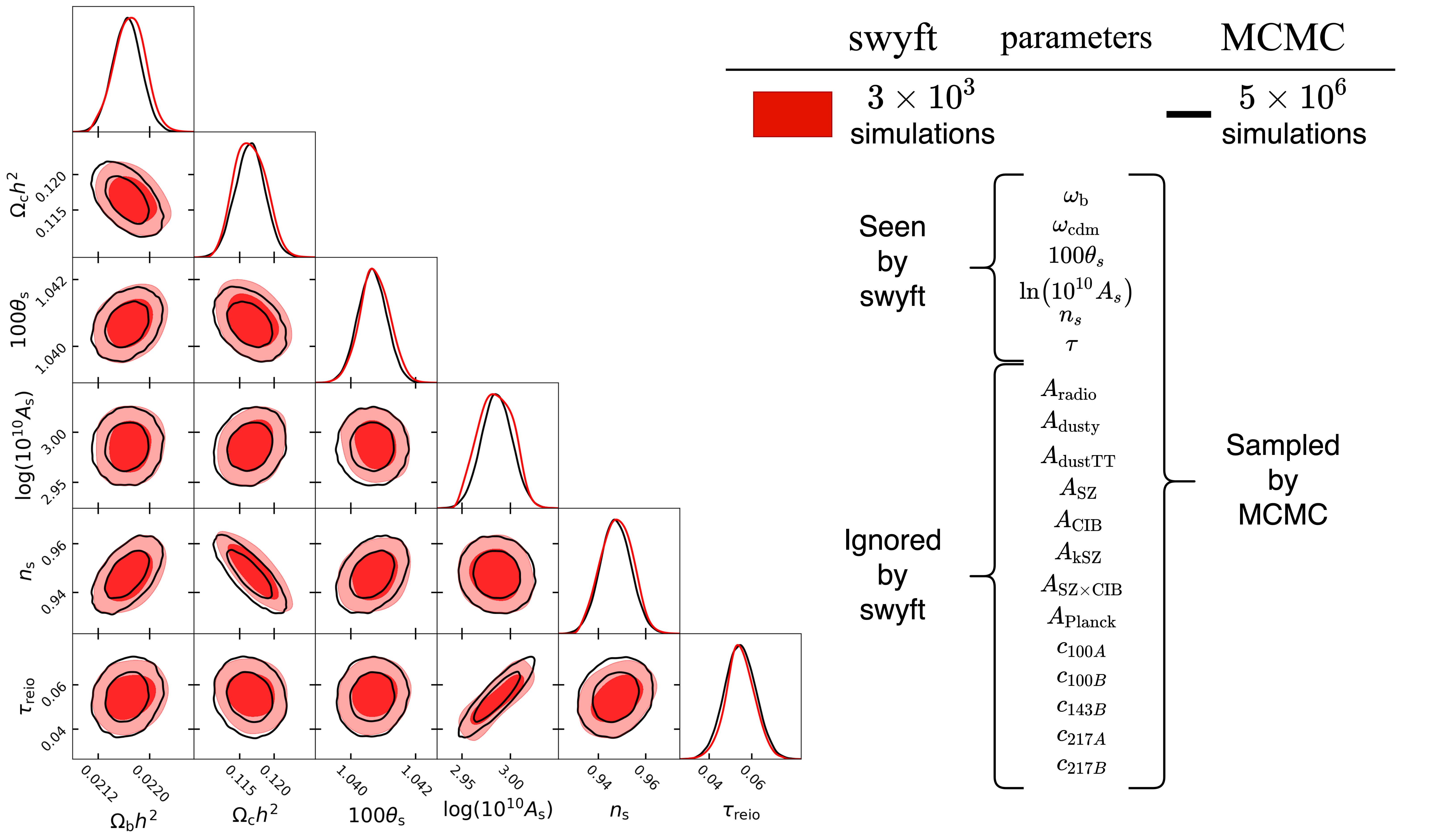}
    \caption{1- and 2-dimensional posteriors inferred by \TMNRE\ trained on $N_{\rm sim}=3000$ realistic simulations, corresponding to the Planck \hillipop\ likelihood. In addition to the 6 cosmological parameters shown, there are 13 varying nuisance parameters. While \TMNRE\ ignores irrelevant parameters when learning posteriors, \MCMC\ must sample each parameter. Converged $2\sigma$-contours for \MCMC\ require $\mathcal{O}(5\times 10^6)$ simulations.}
    \label{fig:hillipop}
\end{figure}

With this prior, we draw $N_{\rm sim}=3000$ simulations and infer posteriors for 6 cosmological parameters using a linear compression network from 10816 to 10 features. We show posteriors for the cosmological parameters in Fig.\ \ref{fig:hillipop}. With only $N_{\rm sim}=3000$ simulations, \TMNRE\ learns accurate 1- and 2-dimensional marginal posteriors, while \MCMC\ requires $\mathcal{O}(5\times 10^6)$ simulator calls for converged 1- and 2-$\sigma$ contours (cf. footnote \ref{foot:converge}). While the simulator cost for \MCMC\ is significantly larger due to the addition of nuisance parameters, \TMNRE\ directly learns marginal posteriors of interest and is not afflicted by the curse of dimensionality.

\subsection{Neutrino masses}\label{sec:global}
As a final application, we turn to neutrino masses, which provide a cosmologically-relevant instance of significantly non-Gaussian posteriors. We use this example to show that for \TMNRE:

\begin{center}
    \textbf{Non-Gaussian posteriors are a non-problem.}
\end{center}

The existence of a cosmological background of relic neutrinos is a standard prediction of the $\Lambda$CDM cosmology. Observations of neutrino oscillations require at least two of the three neutrinos to be massive (with the sum of the three masses $\sum m_\nu \gtrsim 0.06$ eV)~\cite{Esteban:2018azc,deSalas:2020pgw,Capozzi:2017ipn} -- as such, a majority of the energy density stored today in the cosmic neutrino background (CNB) is expected to be in the form of non-relativistic neutrinos. The transition from relativisitic to non-relativistic neutrinos modifies the characteristic free-streaming length, the physical effect of which is to damp the growth of small scale perturbations. From an observational perspective, the presence of neutrino masses appears as a suppression of the matter power spectrum on scales $k \gtrsim k_{nr}$, where $k_{nr} \sim 0.018 \, \Omega_m \, \sqrt{m / {\rm 1 \, eV}} \, h \, {\rm Mpc}^{-1}$ is related to the comoving scale at which the transition occurs (see e.g.~\cite{lesgourgues2006massive,wong2011neutrino,lesgourgues2012neutrino,lesgourgues2014neutrino}). 

At first order, cosmology is only sensitive to the net sum of neutrino masses, placing constraints today at the level of $\sum m_\nu \lesssim 0.12$ eV~\cite{Planck:2018vyg} (to be compared with terrestrial experiments, which currently have constrained $\sum m_\nu \lesssim 0.9$ eV~\cite{aker2022direct}). Terrestrial experiments\footnote{The best terrestrial sensitivity to the sum of neutrino masses comes from tritium beta decay experiments such as KATRIN~\cite{KATRIN:2019yun}.} on the other hand, are typically most sensitive to the relative differences in neutrino masses, their mixings, and possible CP violating phases~\cite{Esteban:2018azc,deSalas:2020pgw,Capozzi:2017ipn}; this implies that cosmological probes of the small scale structure of the universe offer a highly complementary route to testing the properties of neutrinos. Moreover, it seems likely that cosmology will become sensitive to the lower mass limit (\emph{i.e.~} $\sum m_\nu \sim 0.06$ eV) long before their terrestrial counterpart~\cite{wong2011neutrino} (although it is worth emphasizing that the cosmological measurments are sensitive to exotic physics, particularly those that can modify the neutrino phase space distribution). Given the importance of a cosmological neutrino mass measurement, we investigate a final application of the methods outlined here by forecasting the sensitivity of future experiments. This in particular demonstrates \TMNRE's ability to estimate non-Gaussian posteriors. (For the demonstration of this algorithm on more exotic posteriors, see \cite{Miller:2020hua,Miller:2021hys}.)

\begin{figure}[ht]
    \centering
    \includegraphics[width=\textwidth]{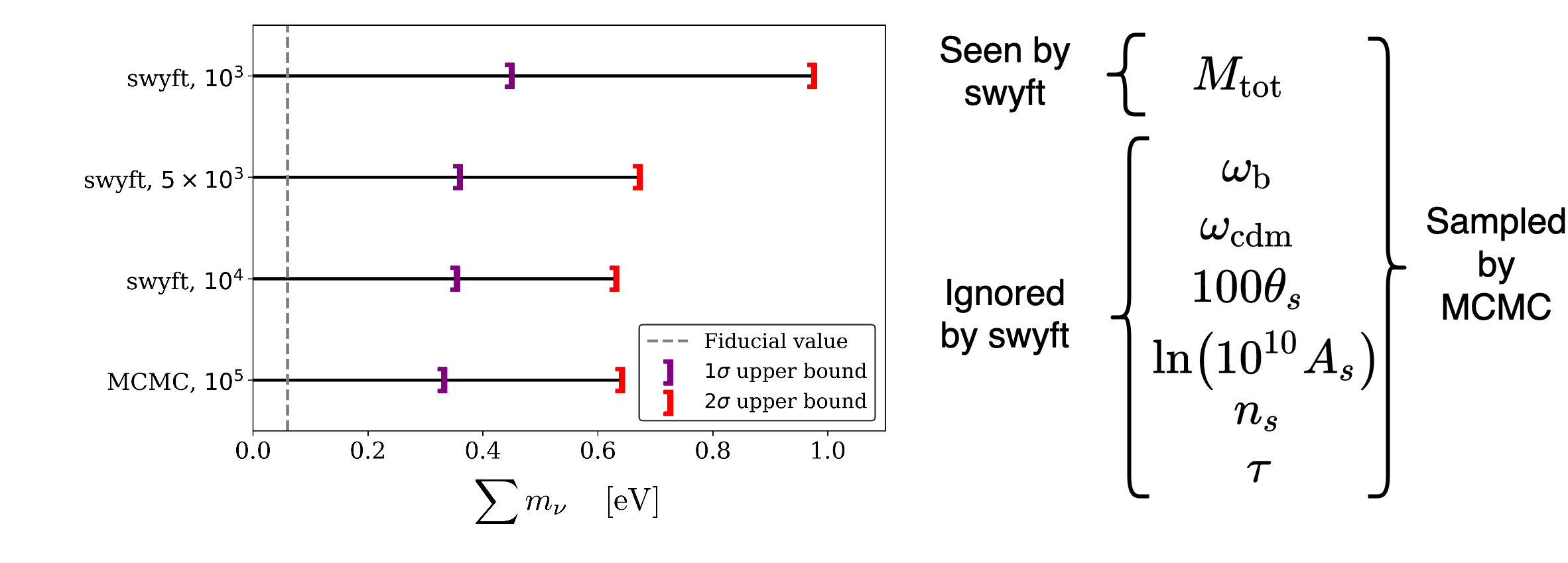}
    \caption{Upper bounds on $M_{\rm tot}$ as estimated by \TMNRE\ for various simulator cost. We see that already at $N_{\rm sim}=5000$, \TMNRE\ accurately approximates the long-run ($10^5$ simulations) \MCMC\ result. \TMNRE\ directly learns the marginal posterior for $M_{\rm tot}$ and thus circumvents the sampling cost associated with $\Lambda$CDM parameters.}
    \label{fig:neutrino}
\end{figure}

Here we consider the simulator of Sec.\ \ref{sec:cltt-warmup}, with the cosmology extended to include massive neutrinos. We include three neutrinos degenerate in mass, each with mass $m_\nu=M_{\rm tot}/3$. Results from the inference are shown in Fig. \ref{fig:neutrino}. 

In particular, we show the 1- and 2-$\sigma$ upper bounds for $M_{\rm tot}$ as computed by \TMNRE\ with various simulator budget. 
As our fiducial cosmology we use the same parameters as in Sec.\ \ref{sec:cltt-warmup} and $M_{\rm tot}=0.06~\textrm{eV}$. For \TMNRE\ we use $N_{\rm sim}\leq 10^4$ simulations drawn from a prior determined, as in Sec.\ \ref{sec:cltt-warmup} using a Fisher matrix formalism and a uniform distribution over $\pm 5\sigma$ ranges. It is well-known (see e.g.\ \cite{Perotto:2006rj}) that the Fisher matrix gives a conservative (i.e.\ too wide) estimate of the posteriors in this case. Regardless, \TMNRE\ produces accurate posteriors with orders of magnitude fewer simulator calls than the ground-truth \MCMC, which requires $\mathcal{O}(10^5)$ simulations for convergence (cf. footnote \ref{foot:converge}). Already at $N_{\rm sim}=5000$, the $2\sigma$ upper bound computed by \TMNRE\ agrees with the long-run \MCMC\ result to within $3\%$.

\subsection{Testing statistical consistency}\label{sec:amort}

So far we have demonstrated the accuracy of \TMNRE\ results by comparing against long-run \MCMC\ chains. Since one of our points is that \TMNRE\ avoids the simulator cost associated with \MCMC, we would like a way to test the consistency of \TMNRE\ results without referring to \MCMC. This is especially important for inferring parameters when a likelihood is not available and the ground-truth is unknown. We now show how consistency checks that are inaccessible to sampling-based methods are possible with \TMNRE.

\begin{center}
    \textbf{Statistical under/overconfidence can be cheaply assessed using local amortization.}
\end{center}

\begin{figure}
    \centering
    \includegraphics[width=\textwidth]{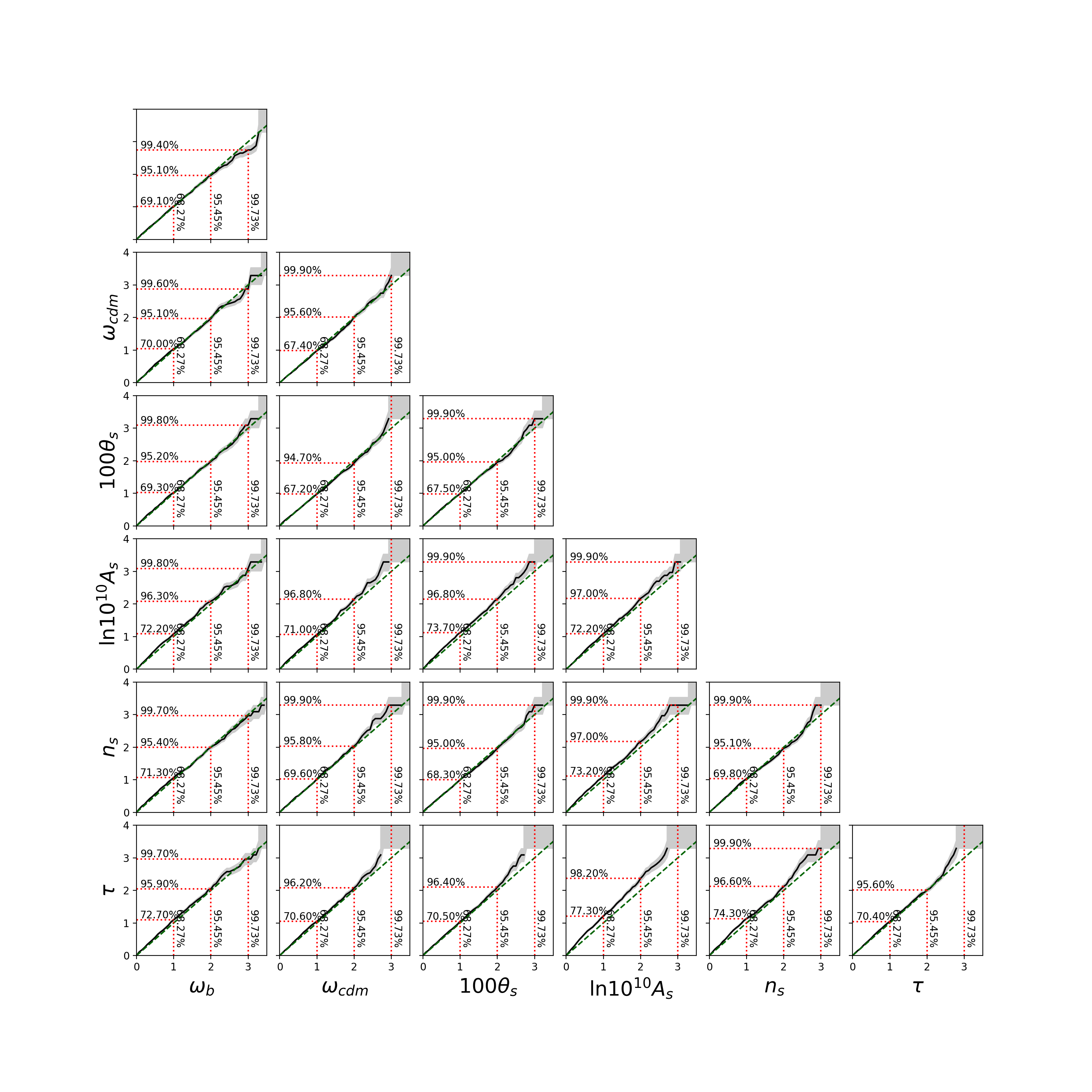}
    \caption{Local amortization allows for cheap consistency checks. Here we show the results of our coverage test applied to all 1- and 2-dimensional posteriors for the network trained in Sec.\ \ref{sec:cltt-warmup}, for which the posteriors are shown in Fig. \ref{fig:CMBForecasting}. }
    \label{fig:amortizationContainment}
\end{figure}

In particular, we perform the test defined in Sec.\ \ref{sec:coverage} for the marginal posteriors trained in Sec.\ \ref{sec:cltt-warmup}.
We show the results in Fig.\ \ref{fig:amortizationContainment}. We see that the empirical coverage and confidence level generally match to excellent precision. Even if they did not match, this information could be used to modify reported constraints compared to network predictions. This test makes no reference to the true parameter values of the observed data or long-run \MCMC\ results, and therefore will prove very important in the inference of observational data with unknown true parameters and implicit likelihood.

\subsection{Wide priors}\label{sec:wideprior}
In the previous sections we have dealt with reasonably narrow priors, estimated either via a Fisher matrix formalism or related observational constraints. However, there are some situations in which the prior is much wider than the support of the posterior, for example in the context of bias parameters in large-scale structure \cite{Oddo:2021iwq}. In this section we demonstrate how wide priors can be accommodated by using multiple rounds of training and truncation with \TMNRE:

\begin{center}
    \textbf{Truncation efficiently identifies the parameter region of interest.}
\end{center}

\begin{figure}
    \centering
    \includegraphics[width=\textwidth]{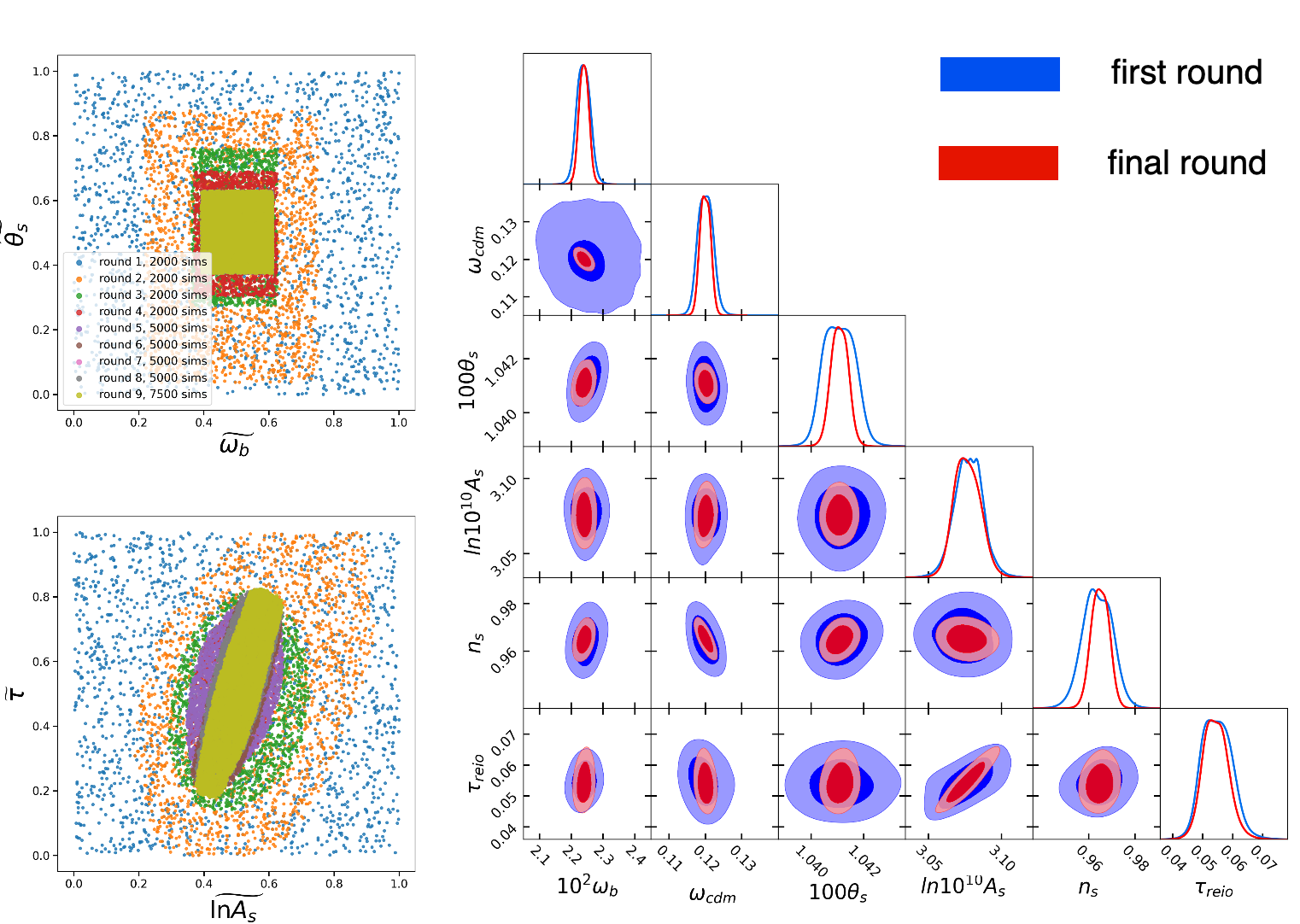}
    \caption{Left: evolution of training data over 9 rounds. Parameters are shown scaled to the unit hypercube. Using expected correlations from the Fisher matrix, $\ln 10^{10}A_s$ and $\tau$ are truncated with correlation to each other, as are $\omega_{\rm cdm}$ and $n_s$ (not shown). $\omega_{\rm b}$ and $100\theta_s$ are truncated independently. Whenever the volume of the truncated prior remains roughly constant between rounds, the amount of training data is increased. Right: comparison of posteriors from the first and last rounds.}
    \label{fig:widePriors}
\end{figure}

Examples outside the cosmological context can be found in \cite{Miller:2020hua,Miller:2021hys}.
As a concrete test, we consider the CMB forecasting simulator from Sec.\ \ref{sec:cltt-warmup}. This time, each cosmological parameter is drawn from a uniform prior with $\theta_i\in [\theta_i^0-25\sigma_F^i, \theta_i^0+25\sigma_F^i]$ with $\sigma_F$ the Fisher estimate, except for $\tau$, for which $\pm 8 \sigma_F$ is used to stay within the physical range. The volume of this prior is 5000 times larger than that in Sec.\ \ref{sec:cltt-warmup}. We perform multiple rounds of training with the same settings as in Sec.\ \ref{sec:cltt-warmup}. When truncating, we take into account correlations for the parameter pairs $(\omega_{\rm cdm},n_s)$ and $\left(\ln\left(10^{10} A_s\right),\tau\right)$ respectively, with no correlations considered for $\omega_b$ and $100\theta_s$. Starting from 2000 simulations in the first round, the amount of training data was increased every time the volume of the truncated prior shrunk by a factor of less than $5$. Over 9 rounds, 20,000 simulations are used, a factor of 4 increase from the inference with ``reasonable'' prior in Sec.\ \ref{sec:cltt-warmup}. Some simulations are reused between rounds.\footnote{We ran this experiment with 10,000 simulations in the initial round, and convergence was also achieved at a total simulation cost of roughly 20,000, in this case distributed over three rounds of training.} The evolution of the training data as well as posteriors from the first and last rounds are shown in Fig. \ref{fig:widePriors}. After the sixth round of training (with $10^4$ simulations seen by the network), the volume of the truncated prior has shrunk by a factor of 5000. In other words, truncation efficiently identifies the parameter region of interest.

\section{Conclusions and outlook}\label{sec:conclusion}
In this work we have illustrated how a novel technique in simulation-based inference (\SBI) called Truncated Marginal Neural Ratio Estimation (\TMNRE, as implemented in the open-source code \swyft\footnote{ \href{https://github.com/undark-lab/swyft}{https://github.com/undark-lab/swyft}}) can be used to improve cosmological parameter inference. In particular, we have argued that \TMNRE\ has several advantages over conventional sampling-based approaches such as \MCMC, including:
\begin{itemize}
    \item {\bf Efficiency.} \TMNRE\ generically requires significantly fewer simulator calls than traditional sampling-based methods simply because it targets the 1- and 2-dimensional marginal posteriors rather than the full joint posterior. In addition, one can alter the priors, the observables, or the inference network without having to re-run the simulator from scratch; in Sec.\ \ref{sec:CMB+BAO}, we illustrated that adding BAO data to the network that was generated using data from a Planck-like simulator required {\emph{zero}} additional calls to the simulator.
    \item {\bf Scalability.} Nuisance parameters are commonplace in cosmology, and tend to significantly impede the convergence timescale of \MCMC. \TMNRE{ is designed in part to soften this scaling with respect to the number of} nuisance parameters introduced; {we present here several examples where parameters} ``ignored'' by the network do not increase the number of simulator calls required for accurate and converged posteriors. This is one of the major advantages offered by \TMNRE\ over other \SBI\ approaches.
    \item {\bf Trustability.} One the most attractive features of \TMNRE\ is that the reliability of the network can be directly tested via local amortization, which effectively amounts to a rigorous check of the coverage of the Bayesian credible interval. This type of test is crucial for establishing that \TMNRE\ is trustable, offering a major advantage over \MCMC\ and some other forms of \SBI.
\end{itemize}

We have illustrated the power of this method using the CMB power spectra and BAO, showing \TMNRE\ generates perfect Bayesian confidence intervals for the $\Lambda$CDM parameters with orders of magnitude less computation time than traditional algorithms. Despite the fact that \TMNRE\ (and more generally \SBI) {provide their biggest improvements} when the likelihood is unknown and simulations are computationally expensive, we have shown that these methods can be powerful even when in circumstances that are more favorable for conventional methods. In particular, we believe that \TMNRE\ offers cosmologists a novel computational tool that mitigates the need for approximation schemes in extended cosmologies (such as those involving massive neutrinos, interacting species, and non-standard distribution functions).

These aspects bode well for extracting cosmological information from future surveys. While in this paper we have worked with simulators for CMB summary statistics, recent work has shown that density-based \SBI\ performs {well} when presented with field-level data \cite{Makinen:2021nly,Villaescusa-Navarro:2021pkb,Villaescusa-Navarro:2021cni} {(and see \cite{Leclercq:2021ctr} for a comparison with field-level likelihood methods). We note that neural ratio estimation, which scales favorably compared to density estimation, has not yet been applied to field-level analysis.} There is no theoretical obstruction to using \SBI\ for full high-resolution simulations; such an approach will likely be crucial to using future data to understand our universe.

To make contact with observational data, it is therefore of high importance for future cosmology experiments to release simulators (which incorporate \eg\ systematic effects) which are capable of reproducing their mock data. Importantly, this does not necessitate a major philosophical shift, since most experiments already release their likelihoods. Moreover, experimental collaborations have already done the hard work of calibrating and validating their simulators. Moving to realistic simulators will not only allow for significant progress in the field of \SBI, but it will also improve the extent to which we can extract cosmological information and build cosmological models.

\subsection*{Acknowledgments and code availability}
We thank Justin Alsing and Ben Wandelt for very useful discussions about the comparison between \pydelfi\ and \TMNRE, and Justin Alsing, Matteo Biagetti, Sascha Caron, Gilles Louppe and Ben Wandelt for comments on a draft of this paper. 
Some of the computations shown were performed using TitanX GPUs on The Distributed ASCI Supercomputer 5 (DAS-5) \cite{7469992}. DAS-5 is funded by the NWO/NCF (the Netherlands Organization for Scientific Research). The project has been partially funded by the Netherlands eScience Center and SURF under grant number ETEC.2019.018.
This work is part of a project that has received funding from the European Research Council (ERC) under the European Union’s Horizon 2020 research and innovation programme (Grant agreement No. 864035).

Software used in this work includes \texttt{numpy} \cite{harris2020array}, \texttt{scipy} \cite{2020SciPy-NMeth}, \texttt{matplotlib} \cite{Hunter:2007}, \texttt{pytorch} \cite{pytorch}, and \texttt{jupyter} \cite{jupyter}.

Accompanying code for the examples in this paper can be found at \href{https://github.com/a-e-cole/swyft-CMB}{https://github.com/a-e-cole/swyft-CMB}.

\appendix
\section{Explicit sky mask}\label{app:mask}
In Sec.\ \ref{sec:cltt-warmup} we incorporated partial sky coverage by modifying the effective number of degrees of freedom with $f_{\rm sky}$. In this appendix, we incorporate an explicit sky mask, which generally makes a likelihood-based analysis significantly more complicated, since different modes are coupled (see e.g.\ \cite{Planck:2019nip}). On the other hand, in the context of \SBI\ incorporating an explicit sky mask is straightforward: we may simply replace masked pixels with some fixed value and compute the corresponding maximum likelihood power spectrum. Here we set masked pixels to $0$. Note that in this context the power spectrum is no longer a sufficient statistic. However, treating CMB analysis at the pixel level at sufficiently high-resolutions will require technical progress beyond the scope of this work.

As a toy example, we infer cosmology using the $TT$ power spectrum mimicking a single frequency channel with WMAP systematics. Namely, we compute the power spectrum up to $\ell_{\rm max}=1024$, employ a Gaussian beam with $FWHM = 0.008$ radians and white noise with $\sigma_T = 35~\mu{\rm K}$. We mask the data using the WMAP 7-year temperature mask\footnote{\href{http://lambda.gsfc.nasa.gov/data/map/dr4/ancillary/masks/wmap_temperature\_analysis\_mask\_r9\_7yr\_v4.fits}{http://lambda.gsfc.nasa.gov/data/map/dr4/ancillary/masks/wmap\_temperature\_analysis\_mask\_r9\_7yr\_v4.fits}}, setting these pixel values to 0, and compute the corresponding maximum likelihood power spectrum.

\begin{figure}
\centering
\includegraphics[width=\textwidth]{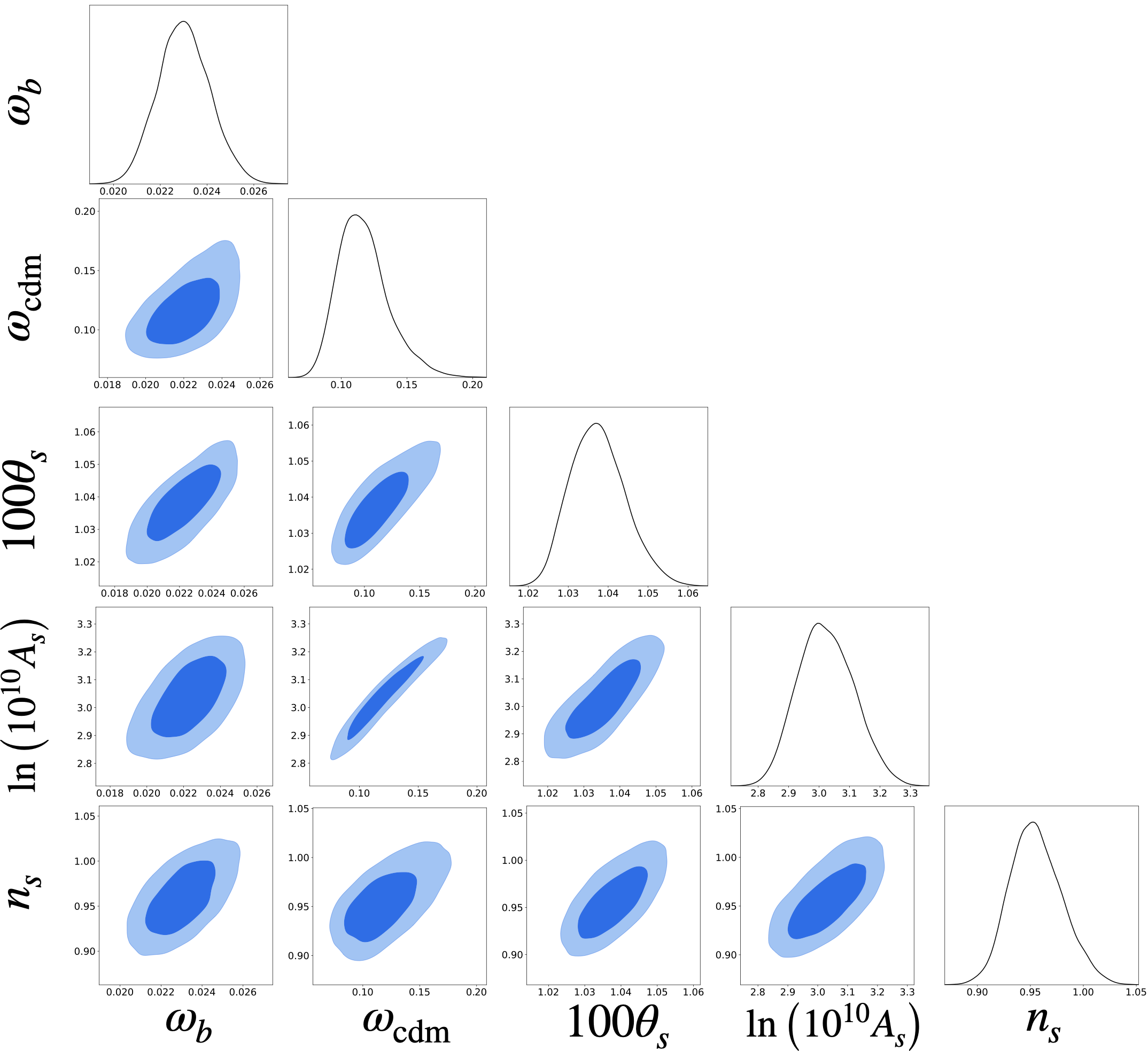}
\caption{One- and two-dimensional posteriors for the power spectrum of an explicitly masked sky.}\label{fig:mask}
\end{figure}

For this simulator, we infer $\Lambda$CDM parameters with $\tau=0.054$ fixed. We take as our prior for the remaining parameters
\begin{align*}
	&\omega_b \sim \mathcal{U}[0.0116, 0.0332]\\
	&\omega_{\rm cdm} \sim \mathcal{U}[0.0206, 0.2193]\\
	&100\theta_s  \sim \mathcal{U}[1.017, 1.065]\\
	&\ln\left(10^{10} A_s\right) \sim \mathcal{U}[2.469, 3.681]\\
	&n_s \sim \mathcal{U}[0.687, 1.243]
\end{align*}
which is wide for all parameters except for $100\theta_s$. Fiducial parameters are the same as those in Sec.\ \ref{sec:cltt-warmup}. We draw 20,000 simulations from this prior and train the 1- and 2-dimensional posteriors, employing a linear embedding layer that compresses the data from 1024 to 10 features. The results are shown in Fig.\ \ref{fig:mask}.

\bibliographystyle{JHEP}
\bibliography{swyftCMB}

\end{document}